\documentstyle[12pt,axodraw]{article}

\newcommand{\mysection}{\setcounter{equation}{0}\section}

\begin{document}
\vskip 0.2cm
\hfill{YITP-SB-04-65}
\vskip 0.2cm
\centerline{\large\bf {An approximation for NLO single Higgs boson inclusive}}
\centerline{\large\bf {transverse momentum distributions in hadron-hadron 
collisions}}
\vskip 0.4cm
\centerline {\sc J. Smith 
\footnote{partially supported
by the National Science Foundation grant PHY-0354776.}}
\centerline{\it C.N. Yang Institute for Theoretical Physics,}
\centerline{\it State University of New York at Stony Brook,
New York 11794-3840, USA.}
\vskip 0.2cm
\centerline {\sc W.L. van Neerven} 
\centerline{\it Instituut-Lorentz}
\centerline{\it University of Leiden,}
\centerline{\it PO Box 9506, 2300 RA Leiden,}
\centerline{\it The Netherlands.}
\vskip 0.2cm
\centerline{January 2005}
\vskip 0.2cm
\centerline{\bf Abstract}
\vskip 0.3cm
In the framework of the gluon-gluon fusion process for Higgs boson production
there are two different prescriptions. They are given by the exact process
where the gluons couple via top-quark loops to the Higgs boson and by the 
approximation
where the top-quark mass $m_t$ is taken to infinity. In the latter 
case the coupling
of the gluons to the Higgs boson is described by an effective Lagrangian. 
Both prescriptions have been used 
for the $2 \rightarrow 2$ body reactions to make 
predictions for Higgs boson production at hadron colliders. 
In next-to-leading order only the effective Lagrangian approach has been used
to compute the single particle inclusive distributions. 
The exact computation of the latter has not been done yet 
because the n-dimensional extensions of $2 \rightarrow 3$ processes are not 
calculated and the two-loop virtual corrections are still missing. To 
remedy this we replace wherever possible the Born cross sections in the 
asymptotic top-quark mass limit by their exact 
analogues. These cross sections appear in the soft and virtual gluon
contributions 
to the next-to-leading order distributions. This approximation is
inspired by the fact that soft-plus-virtual gluons constitute the bulk 
of the higher order correction.
Deviations from the asymptotic top-quark mass limit are discussed.


\vfill

\mysection{Introduction}
\newcommand{\be}{\begin{eqnarray}}
\newcommand{\ee}{\end{eqnarray}}

In the past few years many articles have appeared on searches for the Higgs 
boson
and the reactions in which they are produced. One of them is the 
gluon-gluon fusion
process. According to the standard model gluons do not interact directly
with the Higgs boson but the coupling is mediated by a fermion loop. 
Since the coupling
of the Higgs boson to fermions is proportional to the mass of the fermion
the reaction
proceeds mainly via a top-quark loop \cite{ghkd}. The lowest order loop
is a triangle graph and the Higgs boson decay rates into two gluons or 
two photons were already
calculated at the end of the seventies \cite{wil}. The first calculation in
the gluon-gluon-fusion model for the production process was done at the
end of the eighties by \cite{hino}-\cite{bagl}, (see \cite{kauff},\cite{fida}
for later references). 
Reactions like $g+g\rightarrow g +H$, $q+\bar q\rightarrow g +H$ 
and $q+g\rightarrow q +H$ were calculated. 
In particular the first reaction involves a box diagram leading to
complicated dilogarithms already on the Born level. In the early nineties
people succeeded in calculating the next-to-leading order (NLO) corrections
to the total cross section which involved the computation of the two-loop 
triangular graph with an external Higgs boson \cite{gsz}. The calculation 
could be greatly simplified 
by taking the infinite top-quark mass limit. In this limit the gluons
couple directly to the Higgs boson and the Feynman rules are given by
an effective Lagrangian. It turned out that the latter method gives a good
description of the exact calculation \cite{dawson} provided the Higgs boson 
mass $m_H$
and the transverse momentum $p_t$ are smaller than the top-quark mass 
$m_t$ \cite{ehsb}, \cite{bagl}, \cite{fida}. In particularly the total 
cross section receives its main 
contribution
from small $p_t$. If the Higgs mass is not too large ($m_H<2\,m_t$) the
effective Lagrangian gives a good description of the total cross section
so that recently one has also finished the next-to-next-leading order (NNLO)
computation \cite{harland1}-\cite{rasm1}. However at Higgs masses
and transverse momenta equal or larger than the top-quark mass the 
differential cross
sections calculated with the effective Lagrangian method start to deviate
from the exact cross sections. This has been checked on the Born level
in \cite{ehsb}, \cite{bagl}, \cite{fida}. The investigation should now be done
in NLO but we realize that the exact cross sections are not available.
Differential distributions in NLO using the effective Lagrangian
(or the $m_t\rightarrow \infty$ approach) have been calculated in
\cite{fgk}-\cite{fism}. In the same approach the resummation of the
logarithmically enhanced contributions to
$d\sigma/dp_t$  at small $p_t$ have been carried out in \cite{boca}
-\cite{field}. The first landmark calculation to get the full NLO differential 
distribution has been achieved in \cite{dkosz}. In the latter one
has exactly calculated all matrix elements of the $2\rightarrow 3$
processes. These reactions even contain one-loop five-point functions.
However the calculation of the graphs uses the helicity method 
in four dimensions. To compute the single particle inclusive
process we need the matrix elements in n dimensions. Moreover the
two-loop virtual corrections, which are needed to cancel the infrared
and collinear divergences, have not been calculated yet. Therefore we
propose to
make an approximation by replacing all Born contributions in
the infinite top-quark mass limit by their 
exact analogues in the virtual-plus-soft corrections. 
However this is not
sufficient. We have also to demonstrate that the soft-plus-virtual gluon 
approximation gives a good description of the differential cross section.
Using a certain prescription we can show that this is really the case.

Our paper is organized as follows. In Section 2 we present the formulas
for the exact cross sections and their analogues in the infinite
top-quark mass limit. Then we make approximations for the partonic
soft-plus-virtual and the soft-gluon cross sections. Finally we adopt 
a prescription
how to implement these formulae for the hadronic $p_t$ distributions.
In Section 3 we make comparisons between our approximate differential 
distributions and those which are derived in the limit
$m_t\rightarrow \infty$.


\pagestyle{myheadings}  
\mysection{Approximation to the exact differential cross section for Higgs 
production}
The differential process we study is the semi-inclusive reaction with one 
Higgs boson $H$ in the final state
\begin{eqnarray}
\label{eq2.1}
H_1(P_1)+H_2(P_2)\rightarrow H(q)+'X'\,.
\end{eqnarray}
Here $H_1$ and $H_2$ denote the incoming hadrons and $X$ represents an 
inclusive hadronic state. In our study we limit ourselves to $2\rightarrow 2$ 
and $2\rightarrow 3$ partonic subprocesses.
The kinematics of the $2\rightarrow 2$ reaction is
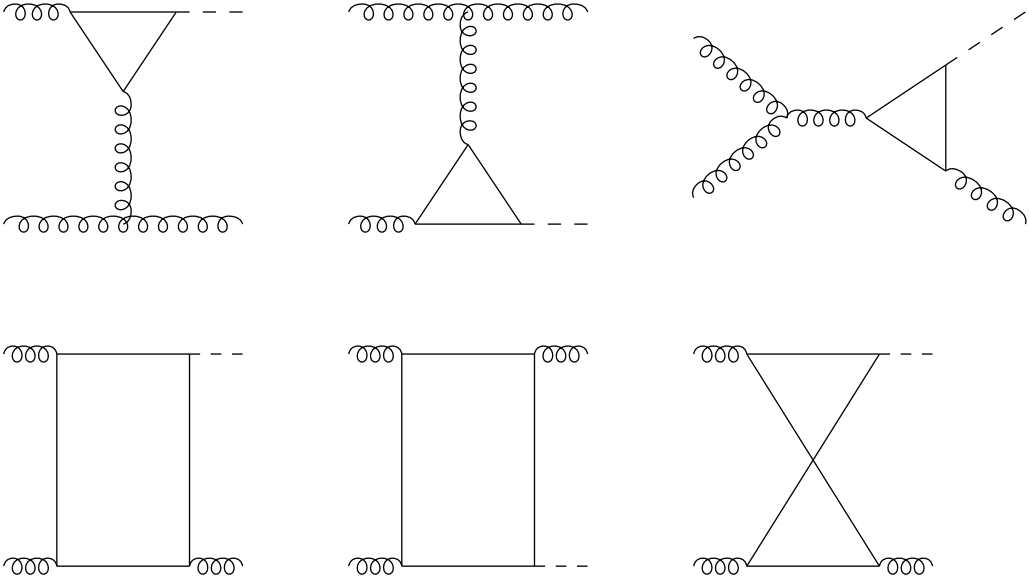
\begin{figure}[ht]
\begin{center}
\begin{picture}(400,100)(0,0)
\Gluon(5,90)(30,90){3}{3}
\DashLine(70,90)(95,90){5}
\Line(30,90)(70,90)
\Line(30,90)(50,60)
\Line(50,60)(70,90)
\Gluon(50,60)(50,10){3}{6}
\Gluon(5,10)(95,10){3}{11}

\Gluon(135,90)(225,90){3}{11}
\Gluon(180,40)(180,90){3}{6}
\Line(160,10)(200,10)
\Line(160,10)(180,40)
\Line(180,40)(200,10)
\Gluon(135,10)(160,10){3}{3}
\DashLine(200,10)(225,10){5}

\Gluon(265,80)(300,50){3}{6}
\Gluon(265,20)(300,50){3}{6}
\Gluon(300,50)(330,50){3}{4}
\Line(330,50)(360,70)
\Line(330,50)(360,30)
\Line(360,70)(360,30)
\Gluon(360,30)(390,10){3}{4}
\DashLine(360,70)(390,90){6}
\end{picture}
\\[10mm]
\begin{picture}(400,100)(0,0)

\Gluon(5,90)(25,90){3}{3}
\DashLine(75,90)(95,90){5}
\Line(25,90)(75,90)
\Line(75,90)(75,10)
\Line(75,10)(25,10)
\Line(25,10)(25,90)
\Gluon(75,10)(95,10){3}{3}
\Gluon(5,10)(25,10){3}{3}

\Gluon(135,90)(155,90){3}{3}
\Gluon(205,90)(225,90){3}{3}
\Line(155,90)(205,90)
\Line(205,90)(205,10)
\Line(205,10)(155,10)
\Line(155,10)(155,90)
\DashLine(205,10)(225,10){5}
\Gluon(135,10)(155,10){3}{3}

\Gluon(265,90)(285,90){3}{3}
\DashLine(335,90)(355,90){5}
\Line(285,90)(335,90)
\Line(335,90)(285,10)
\Line(335,10)(285,10)
\Line(335,10)(285,90)
\Gluon(335,10)(355,10){3}{3}
\Gluon(265,10)(285,10){3}{3}

\end{picture}
\caption[]{ The exact process $g+g \rightarrow g + H$.}
\label{fig1}
\end{center}
\end{figure}

\begin{figure}[ht]
\begin{center}
\begin{picture}(300,100)(0,0)

\DashLine(5,80)(40,50){3}
\DashLine(5,20)(40,50){3}
\Gluon(40,50)(70,50){3}{4}
\Line(70,50)(100,70)
\Line(70,50)(100,30)
\Line(100,70)(100,30)
\Gluon(100,30)(130,10){3}{4}
\DashLine(100,70)(130,90){6}

\Gluon(195,90)(220,90){3}{3}
\DashLine(260,90)(285,90){5}
\Line(220,90)(260,90)
\Line(220,90)(240,60)
\Line(240,60)(260,90)
\Gluon(240,60)(240,10){3}{6}
\DashLine(195,10)(285,10){3}

\end{picture}
\caption[]{ The exact processes $q+\bar q \rightarrow g + H$ and $q(\bar q)+g\rightarrow
q(\bar q) +H$.}
\label{fig2}
\end{center}
\end{figure}
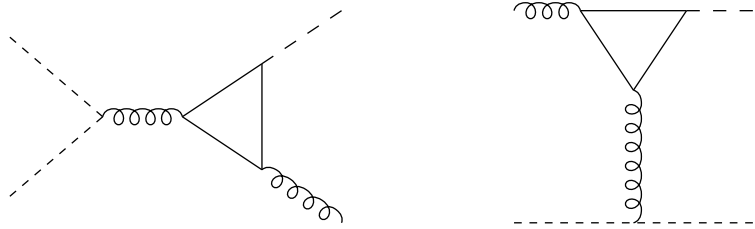
\begin{eqnarray}
\label{eq2.2}
&&a(p_1)+b(p_2) \rightarrow c(p_3) + H(q)\,,
\nonumber\\[2ex]
&& s=(p_1+p_2)^2 \,,\qquad t=(p_1-q)^2 \,,\qquad u=(p_2-q)^2\,.
\end{eqnarray}
The exact calculations of the $2\rightarrow 2$ processes are given in 
\cite{hino}-\cite{fida}. They consist of the following 
parton subprocesses
\begin{eqnarray}
\label{eq2.3}
g+g\rightarrow g+H \,,\qquad q+\bar q\rightarrow g + H \,,\qquad q(\bar q) + 
g\rightarrow q(\bar q) +H\,.
\end{eqnarray}
The Born cross section for the $g+g\rightarrow g+H$ subprocess in 
Fig. \ref{fig1} is equal to
\begin{eqnarray}
\label{eq2.4}
s^2\,\frac{d^2\,\sigma^{(1),{\rm exact}}_{gg\rightarrow g~H}}{dt~du}&=&
\frac{\alpha_w\,\alpha_s^3}{16\,\pi}\,
\frac{1}{s\,t\,u}\,\frac{m_H^8}{M_W^2}\,\frac{N}{N^2-1}\,
\Bigg [|A_2(s,t,u)|^2 +|A_2(u,s,t)|^2
\nonumber\\[2ex]
&&+|A_2(t,u,s)|^2+|A_4(s,t,u)|^2\Bigg ]\,\delta(s+t+u-m_H^2)\,,
\end{eqnarray}
with
\begin{eqnarray}
\label{eq2.5}
 \alpha_w=\frac{e^2}{4\pi\,\sin^2 \theta_W}=\frac{\sqrt{2}\,
M_W^2\,G_F}{\pi}\,,
\end{eqnarray}
where $e$ denotes the electric charge and $\theta_W$ is the weak angle.
The constants $M_W$ and $G_F$ denote the mass of the $W$ and 
the Fermi constant respectively. Further we want to mention that $N=3$ for
QCD.
The dimensionless functions $A_2(s,t,u)$ and $A_4(s,t,u)$ are given in the 
Appendix of
\cite{ehsb}. The Born cross section for the $q+\bar q\rightarrow g+H$ 
subprocess in Fig. \ref{fig2} equals 
\begin{eqnarray}
\label{eq2.6}
s^2\,\frac{d^2\,\sigma^{(1),{\rm exact}}_{q\bar q\rightarrow g~H}}{dt~du}&=&
\frac{\alpha_w\,\alpha_s^3}{128\,\pi}\,
\frac{u^2+t^2}{s\,(u+t)^2}\,\frac{m_H^4}{M_W^2}\,\frac{N^2-1}{N^2}
\,|A_5(s,t,u)|^2
\nonumber\\[2ex]
&&\times \delta(s+t+u-m_H^2)\,,
\end{eqnarray}
where the function $A_5(s,t,u)$ is given in the Appendix of \cite{ehsb}.
Finally the Born cross section for the $q(\bar q)+g\rightarrow q(\bar q)+H$ 
reaction becomes 
(see Fig. \ref{fig2})
\begin{eqnarray}
\label{eq2.7}
s^2\,\frac{d^2\,\sigma^{(1),{\rm exact}}_{qg\rightarrow q~H}}{dt~du}&=&
-\frac{\alpha_w\,\alpha_s^3}{128\,\pi}\,
\frac{u^2+s^2}{t\,(u+s)^2}\,\frac{m_H^4}{M_W^2}\,\frac{1}{N}\,|A_5(t,s,u)|^2
\nonumber\\[2ex]
&&\times \delta(s+t+u-m_H^2)\,.
\end{eqnarray}
In the limit of infinite top mass $m_t$ the functions above simplify enormously.
Actually they can be derived from the effective Lagrangian
\begin{eqnarray}
\label{eq2.8}
{\cal L}_{eff}=G\,\Phi(x)\,O(x)\,, \quad \mbox{with} \quad
O(x)=-\frac{1}{4}\,G_{\mu\nu}^a(x)\,G^{a,\mu\nu}(x)\,,
\end{eqnarray}
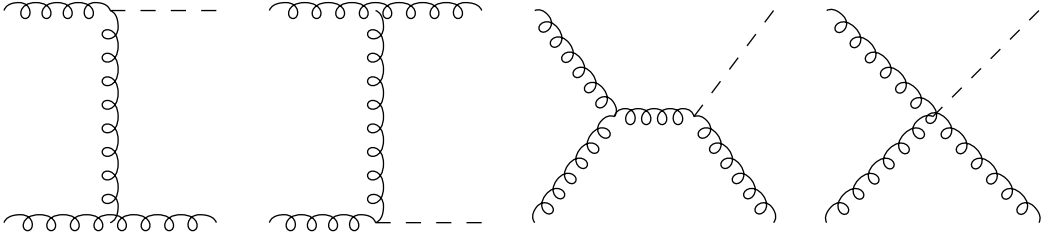
\begin{figure}[t]
\begin{center}
\begin{picture}(400,100)(0,0)

\DashLine(50,90)(90,90){6}
\Gluon(10,90)(50,90){3}{4}
\Gluon(50,90)(50,10){3}{8}
\Gluon(10,10)(90,10){3}{8}

\DashLine(150,10)(190,10){6}
\Gluon(110,90)(190,90){3}{8}
\Gluon(150,90)(150,10){3}{8}
\Gluon(110,10)(150,10){3}{4}

\DashLine(270,50)(300,90){6}
\Gluon(210,90)(240,50){3}{6}
\Gluon(210,10)(240,50){3}{6}
\Gluon(240,50)(270,50){3}{4}
\Gluon(270,50)(300,10){3}{6}

\Gluon(320,90)(400,10){3}{13}
\Gluon(320,10)(360,50){3}{6}
\DashLine(360,50)(400,90){5}

\end{picture}
\caption[]{ The approximate process $g+g \rightarrow g + H$.}
\label{fig3}
\end{center}
\end{figure}
\begin{figure}[t]
\begin{center}
\begin{picture}(300,100)(0,0)

\DashLine(70,50)(100,90){6}
\DashLine(10,90)(40,50){3}
\DashLine(10,10)(40,50){3}
\Gluon(40,50)(70,50){3}{4}
\Gluon(70,50)(100,10){3}{6}

\DashLine(250,90)(290,90){6}
\Gluon(210,90)(250,90){3}{4}
\Gluon(250,90)(250,10){3}{10}
\DashLine(210,10)(290,10){3}

\end{picture}
\caption[]{ The approximate processes $q+\bar q \rightarrow g + H$ and
$q(\bar q)+g\rightarrow q(\bar q) +H$.}
\label{fig4}
\end{center}
\end{figure}
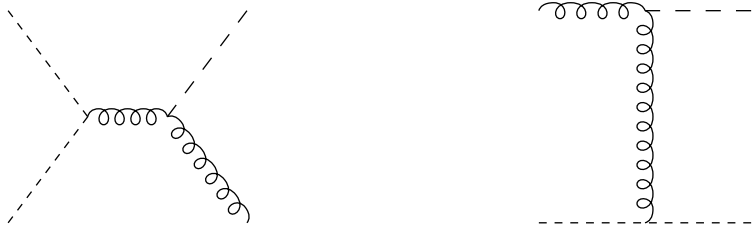
where $\Phi(x)$ represents the Higgs field and $G$ is an effective coupling
constant given by
\begin{eqnarray}
\label{eq2.9}
G^2=\frac{\alpha_w\,\alpha_s^2}{9\,\pi\,M_W^2}\,{\cal C}^2\left
(\alpha_s,\frac{\mu_r^2}{m_t^2}\right )\,.
\end{eqnarray}
The quantity ${\cal C}$ is the coefficient function which describes all QCD 
corrections to the top-quark loops in the limit $m_t\rightarrow \infty$.
For external gluons, which are on-shell, the latter quantity has been
computed up to order $\alpha_s$ in
\cite{gsz}, \cite{dawson}, \cite{daka} and up to $\alpha_s^2$ in
\cite{cks}, \cite{kls}.
Up to second order it reads
\begin{eqnarray}
\label{eq2.10}
{\cal C}\left (\alpha_s(\mu_r^2),\frac{\mu_r^2}{m_t^2}\right )&=&
1+\frac{\alpha_s^{(5)}(\mu_r^2)}{4\pi}\,\Big (11\Big )+
\left (\frac{\alpha_s^{(5)}(\mu_r^2)}{4\pi}\right )^2\,\left [\frac{2777}{18}
+19\,\ln \frac{\mu_r^2}{m_t^2}\right.
\nonumber\\[2ex]
&&\left.+ n_f\,\left (-\frac{67}{6}+\frac{16}{3}\,\ln \frac{\mu_r^2}{m_t^2}
\right ) \right ]\,.
\end{eqnarray}
Here $\mu_r$ represents the renormalization scale and $n_f$ denotes the 
number of light flavours. Moreover
$\alpha_s^{(5)}$ is presented in a five-flavour-number scheme. In the infinite
top-quark mass limit the Feynman rules can be derived from 
Eq. (\ref{eq2.8}). In that limit the Born cross sections become 
\begin{eqnarray}
\label{eq2.11}
s^2\,\frac{d^2\sigma^{(1), m_t\rightarrow \infty}_{gg\rightarrow g~H}}{dt~du}
&=& \frac{\alpha_w\,\alpha_s^3}{144\,\pi}\,\frac{N}{N^2-1}
\,\frac{1}{s\,t\,u\,M_W^2}\,\Bigg [s^4+t^4+u^4+m_H^8\Bigg ]
\nonumber\\[2ex]
&& \times \delta(s+t+u-m_H^2)\,,
\\[2ex]
\label{eq2.12}
s^2\,\frac{d^2\sigma^{(1), m_t\rightarrow \infty}_{q\bar q\rightarrow g~H}}
{dt~du}&=& \frac{\alpha_w\,\alpha_s^3}{288\,\pi}\,
\frac{N^2-1}{N^2}\,\frac{t^2+u^2}{s\,M_W^2}\,\delta(s+t+u-m_H^2)\,,
\\[2ex]
\label{eq2.13}
s^2\,\frac{d^2\sigma^{(1), m_t\rightarrow \infty}_{qg\rightarrow q~H}}{dt~du}
&=& -\frac{\alpha_w\,\alpha_s^3}{288\,\pi}\,\frac{1}{N}
\,\frac{u^2+s^2}{t\,M_W^2}\,\delta(s+t+u-m_H^2)\,,
\end{eqnarray}
where the the graphs are shown in Fig. \ref{fig3} and \ref{fig4}.
The next order gluonic corrections to the $2\rightarrow 2$ reactions
in Figs. \ref{fig1} and \ref{fig2} have not been calculated yet. Some of 
the graphs are shown in Fig. \ref{fig5}, which shows that the calculation
will be very tedious.
However we can make an approximation. In the infinite 
top-quark mass limit the soft-plus-virtual ($S+V$) cross sections could be 
written as \cite{rasm1}
\begin{figure}[t]
\begin{center}
\begin{picture}(400,100)(0,0)

\Gluon(5,90)(30,90){3}{3}
\DashLine(70,90)(95,90){5}
\Line(30,90)(70,90)
\Line(30,90)(50,40)
\Line(50,40)(70,90)
\Gluon(50,40)(50,10){3}{6}
\Gluon(5,10)(95,10){3}{11}
\Gluon(40,65)(60,65){3}{5}

\Gluon(130,90)(225,90){3}{11}
\Gluon(180,40)(180,90){3}{6}
\Line(160,10)(200,10)
\Line(160,10)(180,40)
\Line(180,40)(200,10)
\Gluon(130,10)(160,10){3}{3}
\DashLine(200,10)(225,10){5}
\Gluon(145,10)(180,70){3}{9}

\Gluon(265,90)(285,90){3}{3}
\Gluon(335,90)(355,90){3}{3}
\Line(285,90)(335,90)
\Line(335,90)(335,10)
\Line(335,10)(285,10)
\Line(285,10)(285,90)
\DashLine(335,10)(355,10){5}
\Gluon(265,10)(285,10){3}{3}
\Gluon(285,50)(335,50){3}{8}

\end{picture}
\caption[]{ Samples of two-loop graphs contributing to $g+ g \rightarrow g + H$.}
\label{fig5}
\end{center}
\end{figure}
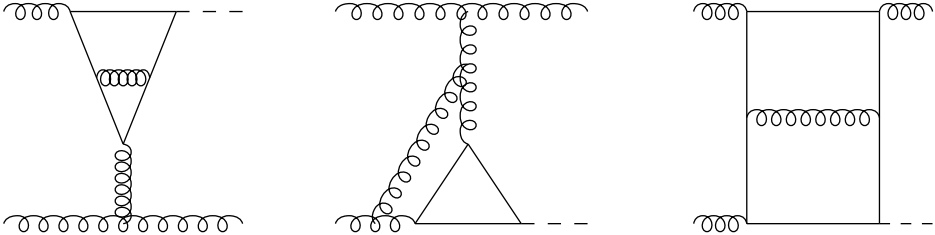
\begin{eqnarray}
\label{eq2.14}
s^2\,\frac{d^2~\sigma^{(2),{\rm S+V}}_{ab\rightarrow c\,H}}{dt~du}&=&
\frac{\alpha_s}{4\pi} \,N(s,t,u,\Delta,\mu^2)\,s^2\,\frac{d^2~\sigma^{(1)}
_{ab\rightarrow c\,H}}{dt~du}
\nonumber\\[2ex]
&&+\pi\,\delta(s+t+u-m_H^2)\,\frac{\alpha_s}{4\pi} \,K\, 
|MB^{(1)}_{ab \rightarrow c\,H}|^2\,.
\end{eqnarray}
Here $d^2\sigma^{(1)}$ denote the Born cross sections in Eqs. (\ref{eq2.11})
-(\ref{eq2.13})
and $MB^{(1)}_{ab \rightarrow c\,H}$
is a left over piece which is numerically very small. The term 
$N(s,t,u,\Delta,\mu^2)$ is an universal function which depends on the parameter
$\Delta$ which serves as a momentum cut off for the infrared divergence. Finally
K denotes a combination of colour factors which vanishes in the supersymmetric
limit $C_A=C_F=n_f=N$. Here $C_A$, $C_F$ are the standard colour factors in
$SU(N)$. For more details see Eqs. (5.24)-(5.26) in \cite{rasm1}. 
Since $N(s,t,u,\Delta,\mu^2)$ is universal we replace the Born cross sections
in the first term of Eq. (\ref{eq2.14}) 
by the exact ones in Eqs. (\ref{eq2.4}), (\ref{eq2.6}), (\ref{eq2.7}). 
In this way we 
get a better soft-plus-virtual gluon approximation for the Higgs boson cross section 
which is also valid for Higgs masses and transverse momenta $p_t$ larger than
the top-quark mass $m_t$. 
\begin{figure}
\begin{center}
\begin{picture}(400,120)(0,0)

\Gluon(10,110)(110,110){3}{11}
\DashLine(80,60)(110,60){5}
\Line(50,40)(50,80)
\Line(50,80)(80,60)
\Line(80,60)(50,40)
\Gluon(50,110)(50,80){3}{6}
\Gluon(50,10)(50,40){3}{6}
\Gluon(10,10)(110,10){3}{11}

\Gluon(130,110)(155,110){3}{5}
\Gluon(205,110)(230,110){3}{5}
\Line(155,110)(205,110)
\Line(205,110)(205,60)
\Line(205,60)(155,60)
\Line(155,60)(155,110)
\DashLine(205,60)(230,60){5}
\Gluon(155,10)(155,60){3}{7}
\Gluon(130,10)(230,10){3}{11}

\Gluon(250,90)(270,90){3}{5}
\Gluon(250,30)(270,30){3}{5}
\Gluon(330,110)(350,110){3}{5}
\Gluon(330,10)(350,10){3}{5}
\DashLine(340,60)(360,60){5}
\Line(270,90)(330,110)
\Line(330,110)(340,60)
\Line(340,60)(330,10)
\Line(330,10)(270,30)
\Line(270,30)(270,90)

\end{picture}
\caption[]{ Samples of graphs contributing to $g+ g \rightarrow g + g + H$.}
\label{fig6}
\end{center}
\end{figure}
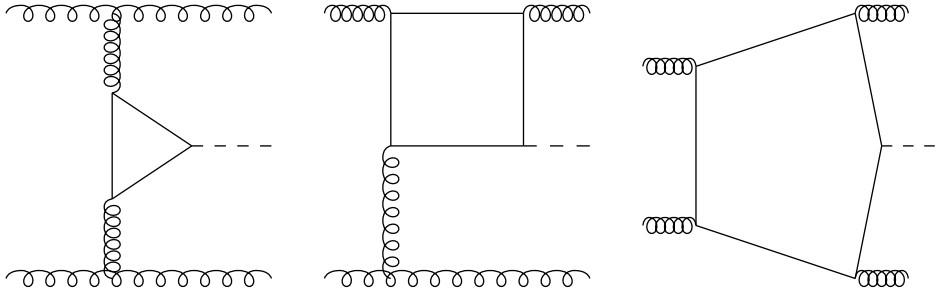
The $2 \rightarrow 3$ reactions are denoted by
\begin{eqnarray}
\label{eq2.15}
&&a(p_1) + b(p_2) \rightarrow c(p_3) + d(p_4) + H(q)\,,
\nonumber\\[2ex]
&&s=(p_1+p_2)^2 \,,\quad t=(p_1-q)^2\,, \quad u=(p_2-q)^2\,, \quad 
s_4=(p_3+p_4)^2\,,
\nonumber\\[2ex]
&&s_4=s+t+u-m_H^2\,.
\end{eqnarray}
The matrix elements for the $2 \rightarrow 3$ processes have been exactly 
calculated 
in \cite{dkosz} although in four dimensions. Some of the graphs are shown 
in Fig. \ref{fig6}.
However we need them in n dimensions to regularize the infrared and 
collinear divergences 
(for $n=4$ there is a problem see \cite{smne}).
Furthermore we also need the exact virtual corrections to cancel the 
infrared divergences.
Since the latter are not calculated yet we can only make an approximation 
for the soft parts 
($s_4 \rightarrow 0$) of the $2 \rightarrow 3$ processes. 
In the $m_t \rightarrow \infty$
limit these parts are
\begin{eqnarray}
\label{eq2.16}
s^2\,\frac{d^2\,\sigma^{(2),{\rm SOFT}}_{ab\rightarrow cd\,H}}{dt\,du}=
\frac{\alpha_s}{4\pi}\,\frac{1}{s_4}\,I(s,t,u,s_4,\mu^2)\,
\,s^2\,\frac{d^2~\sigma^{(1)}_{ab\rightarrow c\,H}}{dt~du}\,,
\end{eqnarray}
where $d^2\sigma^{(1)}$ are the Born cross sections in the limit 
$m_t\rightarrow \infty$ given in
Eqs. (\ref{eq2.11})-(\ref{eq2.13}). The term $I(s,t,u,s_4,\mu^2)$ is a 
universal factor and contains simple functions
which are proportional to $\ln s_4/\mu^2$. For more details see 
Eqs. (5.16)-(5.20) in \cite{rasm1}. We get a better approximation to the 
exact cross sections
if we replace in Eq. (\ref{eq2.16}) the cross sections in $m_t\rightarrow 
\infty$ limit by the exact ones in 
Eqs. (\ref{eq2.4}), (\ref{eq2.6}),(\ref{eq2.7}). The most optimal 
next-to-leading order (NLO) cross section that one
can achieve is to use the exact lowest order cross sections in 
Eq. (\ref{eq2.4}), (\ref{eq2.6}), (\ref{eq2.7}) 
and in next order to substitute them in Eq. (\ref{eq2.14}) and (\ref{eq2.16}). 
This relies upon the fact that the 
soft-plus-virtual gluon approximation
is a very good substitute for the exact cross section. 
We know from our experience with the cross section in the infinite
top-quark mass limit that this is really
the case. This is revealed by a study of the transverse momentum $p_t$
and the rapidity $y$ distributions in Figs. 13-15 of \cite{rasm1}. Above
$p_t=100~{\rm GeV/c}$ and $m_H \ge 100~{\rm GeV/c^2}$ the soft-plus-virtual
gluon approximation accounts for 80 $\%$ of the cross section. However we 
can do 
even better. This becomes clear if we look at the transverse momentum 
distribution
\begin{eqnarray}
\label{eq2.17}
\frac{d~\sigma^{{\rm H_1H_2}}}{d~p_t}(S,p_t^2,m_H^2)=\sum_{a,b=q,g}
\int_{x_{\rm min}}^{x_{\rm max}} dx\,\tilde \Phi_{ab}^{\rm H_1H_2}(x,\mu^2)\,
\frac{d~\sigma_{ab}}{d~p_t}(xS,p_t^2,m_H^2,\mu^2)\,,
\nonumber\\
\end{eqnarray}
with
\begin{eqnarray}
\label{eq2.18}
x_{\rm min}=\frac{m_H^2+2\,p_t^2+2\,\sqrt{p_t^2\,(p_t^2+m_H^2)}}{S}\,,\qquad
x_{\rm max}=1\,,
\end{eqnarray}
and $\tilde \Phi_{ab}$ denotes the momentum fraction luminosity defined by
\begin{eqnarray}
\label{eq2.19}
\tilde \Phi_{ab}^{\rm H_1H_2}(x,\mu^2)=\int_0^1 dx_1\int_0^1dx_2\,
\delta(x-x_1\,x_2)\,
f_a^{\rm H_1}(x_1,\mu^2)\,f_b^{\rm H_2}(x_2,\mu^2)\,.
\end{eqnarray}
However Eq. (\ref{eq2.17}) can also be cast in the form (see \cite{cafl})
\begin{eqnarray}
\label{eq2.20}
\frac{d~\sigma^{{\rm H_1H_2}}}{d~p_t}(S,p_t^2,m_H^2)&=&\sum_{a,b=q,g} 
\,{x_{\rm min}}\, \int_{x_{\rm min}}^{x_{\rm max}} dx\, \Phi_{ab}^{\rm H_1H_2}
(x,\mu^2)\,\frac{x}{x_{\rm min}}\,
\nonumber\\[2ex]
&&\times \frac{d~\sigma_{ab}}{d~p_t}(x~S,p_t^2,m_H^2,\mu^2)\,,
\nonumber\\
\end{eqnarray}
where $\Phi_{ab}$ is the parton luminosity given by
\begin{eqnarray}
\label{eq2.21}
\Phi_{ab}^{\rm H_1H_2}(x,\mu^2)=x^{-1}\,\tilde \Phi_{ab}^{\rm H_1H_2}(x,\mu^2)
\,.
\end{eqnarray}
If we consider the whole cross section it makes no difference which 
definition we
are using. However if we limit ourselves to the soft-plus-virtual gluon 
approximation and moreover we set $x/x_{\rm min}=1$ in Eq. (\ref{eq2.20}) 
we get a difference. In fact we
enhance the small $x$ region which leads to an improvement of the 
approximation. This is mainly due to the fact that the small $x$ 
gluons dominate the differential distributions as they already did in the 
total cross section
(see \cite{harland2}-\cite{rasm1}). This will be shown in the next section.

\mysection{Differential distributions for the LHC and the TEVATRON}
 
In this section the hadronic differential distributions are presented
for arbitrary Higgs mass $m_H$ and top-quark mass $m_t$. We compare the
results for the NLO differential cross sections in the infinite top-quark
mass limit and in the approximation derived in the previous section, 
which is valid for arbitrary $m_t$.  Since the latter is only defined for the
transverse momentum we will limit ourselves to the $p_t$-distributions.
In this paper we will study Higgs boson production in proton-proton
collisions at LHC ($\sqrt{S}=14.0~{\rm TeV}$) and proton-anti-proton collisions
at the TEVATRON  ($\sqrt{S}=2.0~{\rm TeV}$).  The hadronic cross section 
is obtained from the partonic cross section as follows
\begin{eqnarray}
\label{eq3.1}
S^2 \frac{d^2~\sigma^{{\rm H_1H_2}}}{d~T~d~U}(S,T,U,m_H^2)&=& \sum_{a,b=q,g}
\int_{x_{1,{\rm min}}}^1 \frac{dx_1}{x_1} \int_{x_{2,{\rm min}}}^1
\frac{dx_2}{x_2}\,
f_a^{\rm H_1}(x_1,\mu^2)
\nonumber\\[2ex]
&&\times f_b^{\rm H_2}(x_2,\mu^2)\,s^2
\frac{d^2~\sigma_{ab}}{d~t~d~u} (s,t,u,m_H^2,\mu^2)\,.
\nonumber\\
\end{eqnarray}
In analogy to Eq. (\ref{eq2.2}) the hadronic kinematical variables are
defined by
\begin{eqnarray}
\label{eq3.2}
S=(P_1+P_2)^2 \,, \qquad T=(P_1-q)^2\,, \qquad U=(P_2-q)^2 \,,
\end{eqnarray}
where $P_1$ and $P_2$ denote the momenta of hadrons $H_1$ and $H_2$
respectively (see Eq. (\ref{eq2.1})).
In the case parton $p_1$ emerges from hadron $H_1(P_1)$ and parton
$p_2$ emerges from hadron $H_2(P_2)$ we can establish the following relations
\begin{eqnarray}
\label{eq3.3}
&& p_1=x_1\,P_1\,, \qquad p_2=x_2\,P_2 \,,
\nonumber\\[2ex]
&& s=x_1\,x_2\,S \,, \quad t=x_1(T-m_H^2)+m_H^2 \,, \quad u=x_2(U-m_H^2)+m_H^2
\,,
\nonumber\\[2ex]
&& x_{1,{\rm min}}=\frac{-U}{S+T-m_H^2}\,, \qquad
x_{2,{\rm min}}=\frac{-x_1(T-m_H^2)-m_H^2}{x_1S+U-m_H^2}\,.
\end{eqnarray}
From Eq. (\ref{eq3.1}) one can obtain the $p_t$ and $y$ distributions.
Neglecting the masses of the incoming hadrons we have the following
relations
\begin{eqnarray}
\label{eq3.4}
&&T=m_H^2-\sqrt S\,\sqrt{p_t^2+m_H^2}\,\cosh y+\sqrt S\,\sqrt{p_t^2+m_H^2}
\,\sinh y\,,
\nonumber\\[2ex]
&&U=m_H^2-\sqrt S\,\sqrt{p_t^2+m_H^2}\,\cosh y-\sqrt S\,\sqrt{p_t^2+m_H^2}
\,\sinh y\,,
\end{eqnarray}
so that the cross section becomes
\begin{eqnarray}
\label{eq3.5}
S \frac{d^2~\sigma^{{\rm H_1H_2}}}{d~p_t^2~d~y}(S,p_t^2,y,m_H^2)=
S^2 \frac{d^2~\sigma^{{\rm H_1H_2}}}{d~T~d~U}(S,T,U,m_H^2)\,.
\end{eqnarray}
The kinematical boundaries are
\begin{eqnarray}
\label{eq3.6}
m_H^2-S\le T \le 0 \,, \qquad -S-T+m_H^2\le U \le \frac{S~m_H^2}{T-m_H^2}+m_H^2
\,,
\end{eqnarray}
from which one can derive
\begin{eqnarray}
\label{eq3.7}
&& 0\le p_t^2 \le p^2_{t,{\rm max}}\,, \qquad
- \frac{1}{2}\ln \frac{S}{m_H^2}\le y \le \frac{1}{2}\ln \frac{S}{m_H^2}\,,
\nonumber\\[2ex]
&&\mbox{with} \quad p^2_{t,{\rm max}}=\frac{(S+m_H^2)^2}{4~S~\cosh^2 y}-m_H^2
\,,
\end{eqnarray}
or
\begin{eqnarray}
\label{eq3.8}
&&- y_{{\rm max}}\le y \le y_{{\rm max}}\,,
\qquad 0\le p_t^2 \le \frac{(S-m_H^2)^2}{4~S}\equiv p^2_{T,{\rm max}} \,,
\nonumber\\[2ex]
&&\mbox{with} \quad y_{{\rm max}}=
\frac{1}{2}\ln\frac{1+\sqrt{1-sq}}{1-\sqrt{1-sq}}\,,
\qquad sq=\frac{4~S~(p_t^2+m_H^2)}{(S+m_H^2)^2}\,.
\end{eqnarray}
We can
perform the integral over the rapidity and obtain the transverse momentum
distribution
\begin{eqnarray}
\label{eq3.9}
\frac{d~\sigma^{{\rm H_1H_2}}}{d~p_t}(S,p_t^2,m_H^2)=
\int_{-y_{{\rm max}}}^{y_{{\rm max}}} dy
\,\frac{d^2~\sigma^{{\rm H_1H_2}}}{d~p_t~d~y}
(S,p_t^2,y,m_H^2)\,,
\end{eqnarray}
with $y_{{\rm max}}$ given in Eq. (\ref{eq3.8}).
An alternative way to obtain the distribution above is given
in Eq. (\ref{eq2.17}). We checked that both procedures lead to the same
numerical result.

We define what we mean by leading order (LO) and next-to-leading
order (NLO). In the infinite top-quark mass limit and in the exact computation
the differential cross section in LO is defined by
\begin{eqnarray}
\label{eq3.10}
\frac{d~\sigma^{\rm LO}}{d~p_t}(S,p_t^2,m_H^2)=
\frac{d~\sigma^{(1)}}{d~p_t}(S,p_t^2,m_H^2)\,,
\end{eqnarray}
where we shall denote the LO cross section in the infinite top-quark
mass limit by $d\sigma^{\rm LO, m_t\rightarrow \infty}/dp_t$.
The partonic cross sections in the latter quantity are given in 
Eqs. (\ref{eq2.11})-(\ref{eq2.13}). The exact LO cross section is represented
by $d\sigma^{\rm LO, exact}/dp_t$ 
with the partonic cross sections in Eqs. (\ref{eq2.4}),
(\ref{eq2.6}) and (\ref{eq2.7}). The
gluon-gluon-Higgs coupling is given by $G$ in Eq. (\ref{eq2.9}) with
${\cal C}=1$. The top-quark mass is given by
$m_t=174.3~{\rm GeV/c^2}$ and the Fermi constant
$G_F=1.16639\times 10^{-5}\,{\rm GeV}^{-2}=4541.68~{\rm pb}$ in 
Eq. (\ref{eq2.5}).
We also adopt the leading logarithmic
representation for the running coupling and the parton densities.
For the latter we choose the parametrization according to \cite{mrst1}
(namely set ${\rm lo2002.dat}$) with $\Lambda_5^{\rm LO}=167~{\rm MeV}$ 
and $n_f=5$. 

The NLO corrected differential cross section in the
asymptotic top-quark mass limit is given by
\begin{eqnarray}
\label{eq3.11}
\frac{d~\sigma^{\rm NLO, m_t\rightarrow \infty}}{d~p_t}(S,p_t^2,m_H^2)&=&
\left [1+22\,\left (\frac{\alpha_s^{(5)}(\mu^2)}{4\pi}\right )\right ]\,
\frac{d~\sigma^{\rm LO, m_t\rightarrow \infty}}{d~p_t}(S,p_t^2,m_H^2)
\nonumber\\[2ex]
&&+\frac{d~\sigma^{(2), {\rm m_t\rightarrow \infty}}}{d~p_t}(S,p_t^2,m_H^2)
\,,
\end{eqnarray}
In $d~\sigma^{(2), {\rm m_t\rightarrow \infty}}$ all partonic cross sections 
use the asymptotic top-quark mass limit results in 
\cite{fgk}-\cite{fism}. Further we have multiplied the LO cross section
by ${\cal C}^2=1+22~\alpha_s/4\pi$ in Eq. (\ref{eq2.10}). 
Finally we have the approximation for arbitrary masses $m_H$ and $m_t$
\begin{eqnarray}
\label{eq3.12}
\frac{d~\sigma^{\rm NLO, approx}}{d~p_t}(S,p_t^2,m_H^2)&=&
\frac{d~\sigma^{\rm LO, exact}}{d~p_t}(S,p_t^2,m_H^2)
\nonumber\\[2ex]
&&+\frac{d~\sigma^{\rm S+V, approx}}{d~p_t}(S,p_t^2,m_H^2) \,,
\end{eqnarray}
where the partonic cross sections are given in Eqs. 
(\ref{eq2.4}), (\ref{eq2.6}) and (\ref{eq2.7}) and the soft-plus-virtual
gluon approximation is given in Eqs. (\ref{eq2.14}) and (\ref{eq2.16}).
The running coupling and parton densities are also
represented in next-to-leading order for which we have chosen
the $\overline{\rm MS}$-scheme and $n_f=5$. For our plots
we have adopted the parametrization obtained
from the set MRST \cite{mrst2} (namely set ${\rm alf119.dat}$)
with $\Lambda_5^{\rm NLO}=239~{\rm MeV}$. 
For simplicity the factorization scale $\mu$ is set equal to the
renormalization scale $\mu_r$. For our plots we take
$\mu=\mu_0=\sqrt{p_t^2+m_H^2}$ unless mentioned otherwise.

Our first study concerns the validity of the soft-plus-virtual gluon 
approximation. This is done in the asymptotic top-quark limit where
we know the complete NLO correction. For that purpose we plot
\begin{eqnarray}
\label{eq3.13}
R=\frac{d\sigma^{\rm S+V, m_t\rightarrow \infty}/dp_t}
{d\sigma^{\rm NLO, m_t\rightarrow \infty }/dp_t}
\end{eqnarray}
in the range $40\,{\rm GeV/c}<p_t< 200\,{\rm GeV/c}$ and 
$m_H=120,~160,~200\,{\rm GeV/c^2}$. The plots are given for the LHC
($\sqrt{S}=14~{\rm GeV}$)
in Fig. 7. The figure reveals that at $m_H=120 \,{\rm GeV/c^2}$
and $p_t=40\,{\rm GeV/c}$ the ratio is 
1.06 and it decreases to about 0.9 at $p_t=200\,{\rm GeV/c}$. For larger
Higgs masses the ratio becomes closer to unity at $p_t>100~{\rm GeV/c}$.  
This feature can be understood because at larger Higgs masses the 
kinematics are closer to the boundary of phase space. 
The conclusion is that in the range
$100\,{\rm GeV/c}<p_t< 200\,{\rm GeV/c}$ we have $0.9<R<1.0$ which indicates
that the soft-plus-virtual gluon approximation 
with the prescription in \cite{cafl} works rather well.  
This is mainly due to the dominance of
the $gg$-channel and the steeply rising gluon flux which is even 
enhanced by the definition of the parton luminosity in Eqs. (\ref{eq2.20}), 
(\ref{eq2.21}). In the case of the TEVATRON ($\sqrt{S}=2~{\rm GeV}$) 
the soft-plus-virtual gluon approximation works even better (see Fig. 8).
In the whole range $40\,{\rm GeV/c}<p_t< 200\,{\rm GeV/c}$ we have
$0.95<R<1.07$. However the mass range is more limited i.e.
$m_H=120,~130,~140\,{\rm GeV/c^2}$ because at larger masses the cross
section becomes unobservably small. This is understandable because at lower 
energies we are closer to the boundary of phase space where
the soft-plus-virtual gluon approximation approaches the exact cross section.
The transverse momentum distributions $d~\sigma/d~p_t$ are plotted in the 
case of the LHC 
in Figs. 9, 10, 11 for $m_H=120,~160,~200\,{\rm GeV/c^2}$ respectively.
The figures reveal the differences between the cross sections in the
asymptotic $m_t$ limits and the exact (approximate) cross sections. 
They becomes more clear if we plot the ratios
\begin{eqnarray}
\label{eq3.14}
H^{\rm LO}=\frac{d\sigma^{\rm LO, exact}/dp_t}
{d\sigma^{\rm LO, m_t\rightarrow \infty}/dp_t}\,, \qquad
H^{\rm NLO}=\frac{d\sigma^{\rm NLO, approx}/dp_t}
{d\sigma^{\rm NLO, m_t\rightarrow \infty}/dp_t} \,,
\end{eqnarray}
which are shown in Figs. 12, 13 and 14.
For the Born cross section they vary at $p_t=40~{\rm GeV/c}$
from 0.93  to 1.03 for $m_H=120\,{\rm GeV/c^2}$ to $200\,{\rm GeV/c^2}$ 
respectively. At $p_t=200~{\rm GeV/c}$ they all become about 0.8
irrespective of the Higgs mass.
For the NLO cases these values are 0.96 to 1.28 for small $p_t$ 
and 0.68 to 0.75 at large $p_t$ as $m_H$ increases from 
$m_H=120~{\rm GeV/c^2}$ to $m_H=200~{\rm GeV/c^2}$.
The exact (approximate) cross sections are always below those in the asymptotic 
$m_t$ limit  except for $m_H=160~{\rm GeV/c^2}$ and $m_H=200~{\rm GeV/c^2}$ 
at small $p_t$. There are cross over points at $p_t=55~{\rm GeV/c}$ 
and $p_t=75~{\rm GeV/c}$ for the NLO cross sections. 
The NLO corrections are very large. This becomes clear if we look at 
the K-factors defined by
\begin{eqnarray}
\label{eq3.15}
K=\frac{d~\sigma^{\rm NLO, approx}/d~p_t}
{d~\sigma^{\rm LO, exact}/d~p_t}\,,
\end{eqnarray}
which are shown in Fig. 15. At $p_t=40~{\rm GeV/c}$ the K-factors
vary from 1.66 to 2.02 as $m_H$ increases from $m_H=120~{\rm GeV/c^2}$
to $m_H=200~{\rm GeV/c^2}$ respectively. At larger $p_t$ values 
the K-factors decrease and at $p_t=120~{\rm GeV/c}$ they stabilise 
around the values
1.5, 1.55, 1.65 for $m_H=120, 160, 200\,{\rm GeV/c^2}$ respectively.
For $m_H=120~{\rm GeV/c^2}$ the difference between the asymptotic
$m_t$ limit and the soft-plus-virtual gluon approximation in NLO
is of the same order as the K-factors, namely 1.5.
In Fig. 16 the transverse momentum distributions are shown for the 
TEVATRON at $m_H=120~{\rm GeV/c^2}$ and in Fig. 17 the ratios in 
Eq. (\ref{eq3.14})
are plotted. Here the discrepancies in NLO are even larger. There is 
very little difference between small $p_t$ and large $p_t$ and the 
approximate cross section
is about 0.5 to 0.8 times smaller than the one in the asymptotic $m_t$ limit. 
The Born approximations vary from 0.8 to 1.25 when $p_t$ ranges from
$p_t=40~{\rm GeV/c}$ to $p_t=200~{\rm GeV/c}$. Notice that for 
$p_t>135~{\rm GeV/c}$ the approximate
cross section becomes even a little bit larger than the one in the case of
the asymptotic $m_t$ limit. The K-factors (see Fig. 18) are a little bit 
smaller than
in the case of the LHC. At $p_t=40~{\rm GeV/c}$ they vary between
1.5 and 1.8 and at larger $p_t$ (say $p_t>120~{\rm GeV/c}$) they are in
the range $1.3<K<1.4$. Here the discrepancy between the asymptotic
$m_t$ limit
and the approximate cross section in NLO is even larger
than the corresponding K-factor.
Note that the peaks in Figs. 16 and 17 reflect the thresholds in the
partonic channels described in \cite{hino}-\cite{fida}.
The dependence of the exact Born and the soft-plus-virtual gluon approximation
cross sections on the factorization scale $\mu$ is studied for 
$m_H=120~{\rm GeV/c^2}$ at $p_t=100, 150, 200~{\rm GeV/c}$.
The dependence can be expressed by the following quantity
\begin{eqnarray}
\label{eq3.16}
N\left (p_t,\frac{\mu}{\mu_0}\right )=
\frac{d\sigma^{\rm approx}(p_t,\mu)/dp_t}{d\sigma^{\rm approx}(p_t,\mu_0)/dp_t}
\end{eqnarray}
with $\mu_0=\sqrt{p_t^2+m_H^2}$. This quantity is plotted in the
range $0.1~\mu_0<\mu<10~\mu_0$ for LO and NLO in Fig. 19 for the LHC and in 
Fig. 20 for the TEVATRON both at $m_H=120~{\rm GeV/c^2}$. 
The LO cross sections have the larger values for small $\mu/\mu_0$.
What is very striking is the improvement in 
scale variation while going from LO to NLO. In LO there is steep behaviour
at small $\mu/\mu_0$ which is flattened out in NLO. At large $\mu/\mu_0$ 
the difference between LO and NLO is not so big, but 
still the NLO curves are flatter than
the LO ones. Basically the same curves are also found at larger 
Higgs masses so that there is no need to show them.
Finally there is a small dependence of $N(p_t,\mu/\mu_0)$ on the transverse 
momenta in both LO and in NLO. 

Concluding our findings we observe that the soft-plus-virtual gluon
approximation gives a good description of the exact NLO cross section
(within 90 $\%$), when tested with $m_t \rightarrow \infty$ cross sections. 
The difference between the asymptotic $m_t$ limit and
the soft-plus-virtual gluon approximation is larger than the K-factor
in the case of the TEVATRON but smaller than the K-factor in the case
of the LHC. Also the validity of asymptotic $m_t$ limit depends more
on the value of the transverse momentum than on the magnitude of the
Higgs mass. 
Finally our approximation has a significantly smaller scale dependence 
for both colliders in particular at small factorization scale.

\appendix
%

\centerline{\bf \large{Figure Captions}}
\begin{description}
\item[Fig. 7.]
The quality of the soft-plus-virtual gluon approximation represented
by the ratio $R$ in Eq. (\ref{eq3.13}) for $40\,{\rm GeV/c}<p_t< 
200\,{\rm GeV/c}$ at the LHC ($\sqrt{S}=14~{\rm TeV}$) for 
$m_H=120\,{\rm GeV/c^2}$ (solid line),
$m_H=160\,{\rm GeV/c^2}$ (dashed line), $m_H=200\,{\rm GeV/c^2}$ (dotted line).
\item[Fig. 8.]
The quality of the soft-plus-virtual gluon approximation represented
by the ratio $R$ in Eq. (\ref{eq3.13}) for $40\,{\rm GeV/c} <p_t< 
200\,{\rm GeV/c}$ at the TEVATRON ($\sqrt{S}=2~{\rm TeV}$) for 
$m_H=120\,{\rm GeV/c^2}$ (solid line),
$m_H=130\,{\rm GeV/c^2}$ (dashed line), $m_H=140\,{\rm GeV/c^2}$ (dotted line).
\item[Fig. 9.]
Differential cross sections at the LHC ($\sqrt{S}=14~{\rm TeV}$) with
$m_H=120\,{\rm GeV/c^2}$.
The Born cross sections $d\sigma^{\rm LO, exact}/dp_t$ (dotted line)
and $d\sigma^{\rm LO, m_t\rightarrow \infty}/dp_t$ (dot-dashed line).
Also shown are the NLO contributions $d\sigma^{\rm NLO, approx}/dp_t$
(solid line) and $d\sigma^{\rm NLO, m_t\rightarrow \infty}/dp_t$
(dashed line).
\item[Fig. 10.]
Differential cross sections at the LHC ($\sqrt{S}=14~{\rm TeV}$)
with $m_H=160\,{\rm GeV/c^2}$.
The Born cross sections $d\sigma^{\rm LO, exact}/dp_t$ (dotted line)
and $d\sigma^{\rm LO, m_t\rightarrow \infty}/dp_t$ (dot-dashed line).
Also shown are the NLO contributions $d\sigma^{\rm NLO, approx}/dp_t$
(solid line) and $d\sigma^{\rm NLO, m_t\rightarrow \infty}/dp_t$
(dashed line).
\item[Fig. 11.]
Differential cross sections at the LHC ($\sqrt{S}=14~{\rm TeV}$)
with $m_H=200\,{\rm GeV/c^2}$.
The Born cross sections $d\sigma^{\rm LO, exact}/dp_t$ (dotted line)
and $d\sigma^{\rm LO, m_t\rightarrow \infty}/dp_t$ (dot-dashed line).
Also shown are the NLO contributions $d\sigma^{\rm NLO, approx}/dp_t$
(solid line) and $d\sigma^{\rm NLO, m_t\rightarrow \infty}/dp_t$
(dashed line).
\item[Fig. 12.]
The factors $H^{\rm LO}$ (dashed line) and 
$H^{\rm NLO}$ (solid line) in Eq. (\ref{eq3.14})
at the LHC ($\sqrt{S}=14~{\rm TeV}$) with $m_H=120\,{\rm GeV/c^2}$. 
\item[Fig. 13.]
Same as in Fig. 12 for $m_H=160\,{\rm GeV/c^2}$.
\item[Fig. 14.]
Same as in Fig. 12 for $m_H=200\,{\rm GeV/c^2}$.
\item[Fig. 15.]
The K factor (Eq. (\ref{eq3.15})) for the LHC ($\sqrt{S}=14~{\rm TeV}$).
$m_H=120\,{\rm GeV/c^2}$ (solid line), $m_H=160\,{\rm GeV/c^2}$ (dashed
line), $m_H=200\,{\rm GeV/c^2}$ (dotted line).
\item[Fig. 16.]
Same as in Fig. 9 but then for the TEVATRON ($\sqrt{S}=2~{\rm TeV}$)
and $m_H=120\,{\rm GeV/c^2}$.
\item[Fig. 17.]
Same as in Fig. 12 but then for the TEVATRON ($\sqrt{S}=2~{\rm TeV}$)
and $m_H=120\,{\rm GeV/c^2}$.
\item[Fig. 18.]
The K factor (Eq. (\ref{eq3.15})) for the TEVATRON ($\sqrt{S}=2~{\rm TeV}$).
$m_H=120\,{\rm GeV/c^2}$ (solid line), $m_H=130\,{\rm GeV/c^2}$ (dashed
line), $m_H=140\,{\rm GeV/c^2}$ (dotted line).
\item[Fig. 19.]
The scale dependence represented by $N(p_t,\mu/\mu_0)$ in Eq. (\ref{eq3.16})
for the LHC ($\sqrt{S}=14~{\rm TeV}$) and $m_H=120\,{\rm GeV/c^2}$. The results
are plotted in the range $0.1<\mu/\mu_0<10$ with $\mu_0^2=m_H^2+p_t^2$ 
for $p_t=100\,{\rm GeV/c}$ (solid line), $p_t=150\,{\rm GeV/c}$ (dashed line)
$p_t=200\,{\rm GeV/c}$ (dotted line).
The upper three curves are for $d\sigma^{\rm LO, exact}/dp_t$
whereas the lower three curves are for $d\sigma^{\rm NLO, approx}/dp_t$.
\item[Fig. 20.]
The scale dependence represented by $N(p_t,\mu/\mu_0)$ in Eq. (\ref{eq3.16})
for the TEVATRON ($\sqrt{S}=2~{\rm TeV}$) and $m_H=120\,{\rm GeV/c^2}$. 
The results
are plotted in the range $0.1<\mu/\mu_0<10$ with $\mu_0^2=m_H^2+p_t^2$
for $p_t=100\,{\rm GeV/c}$ (solid line), $p_t=150\,{\rm GeV/c}$ (dashed line)
$p_t=200\,{\rm GeV/c}$ (dotted line).
The upper three curves are for $d\sigma^{\rm LO, exact}/dp_t$
whereas the lower three curves are for $d\sigma^{\rm NLO, approx}/dp_t$.
\end{description}

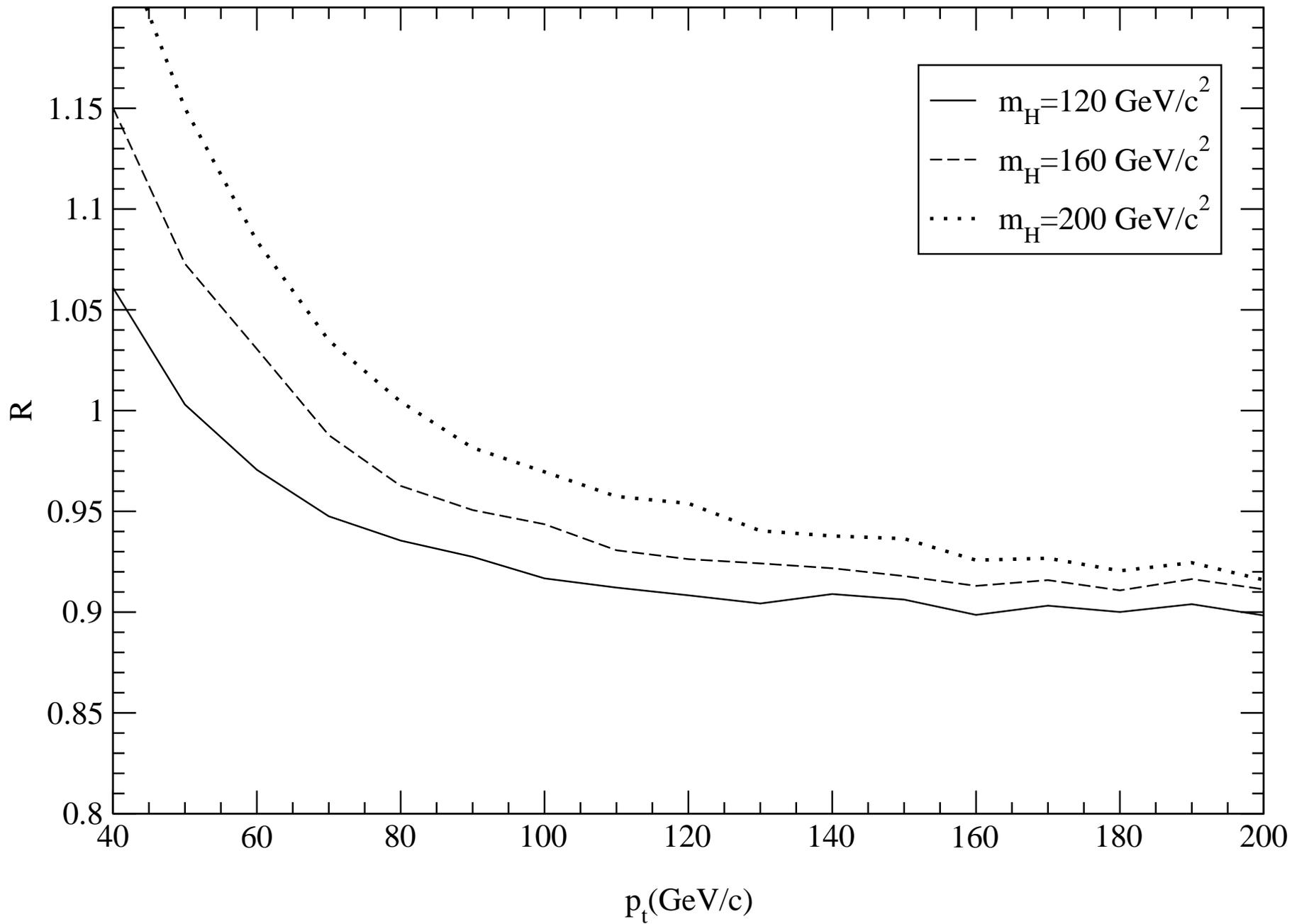
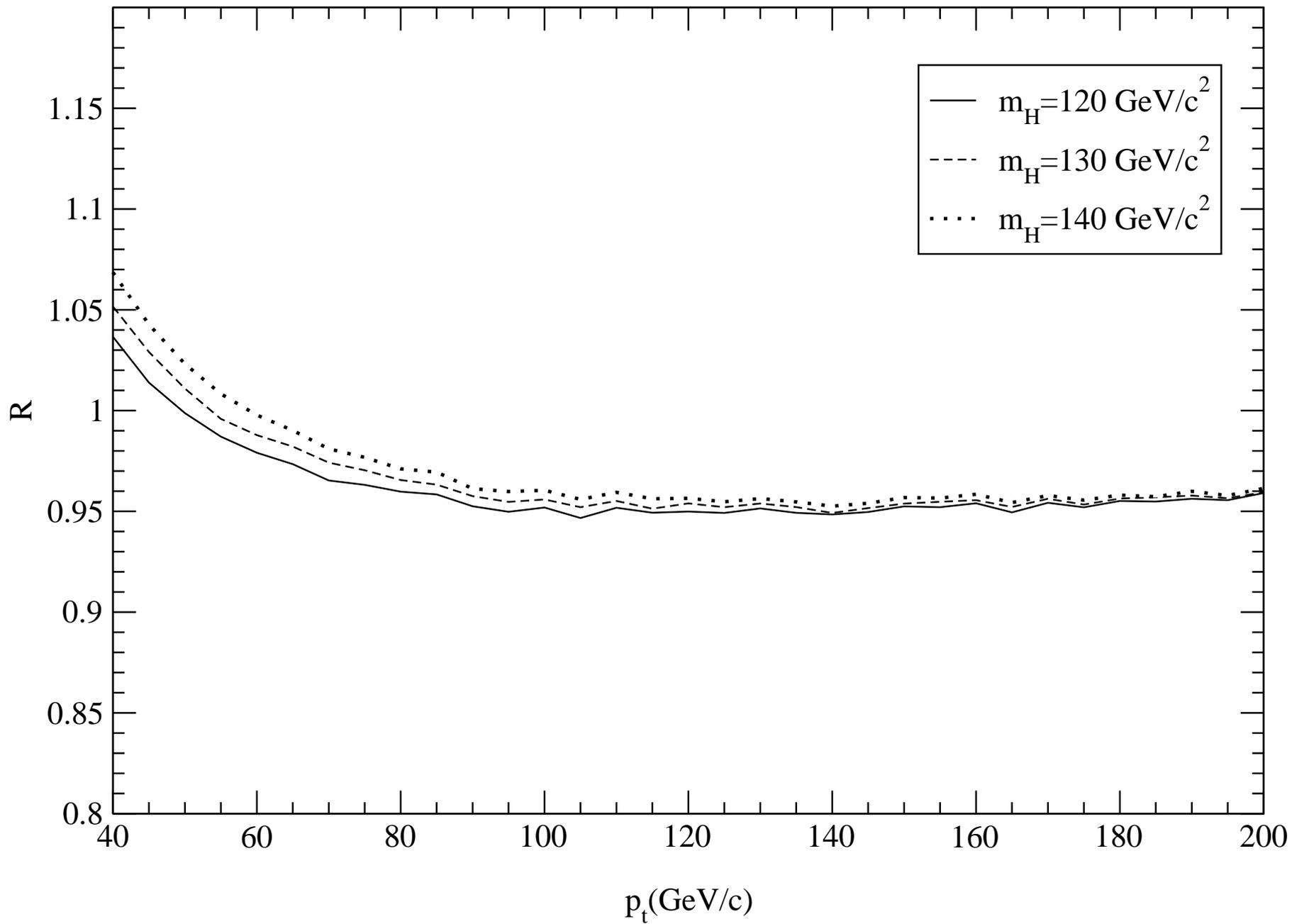
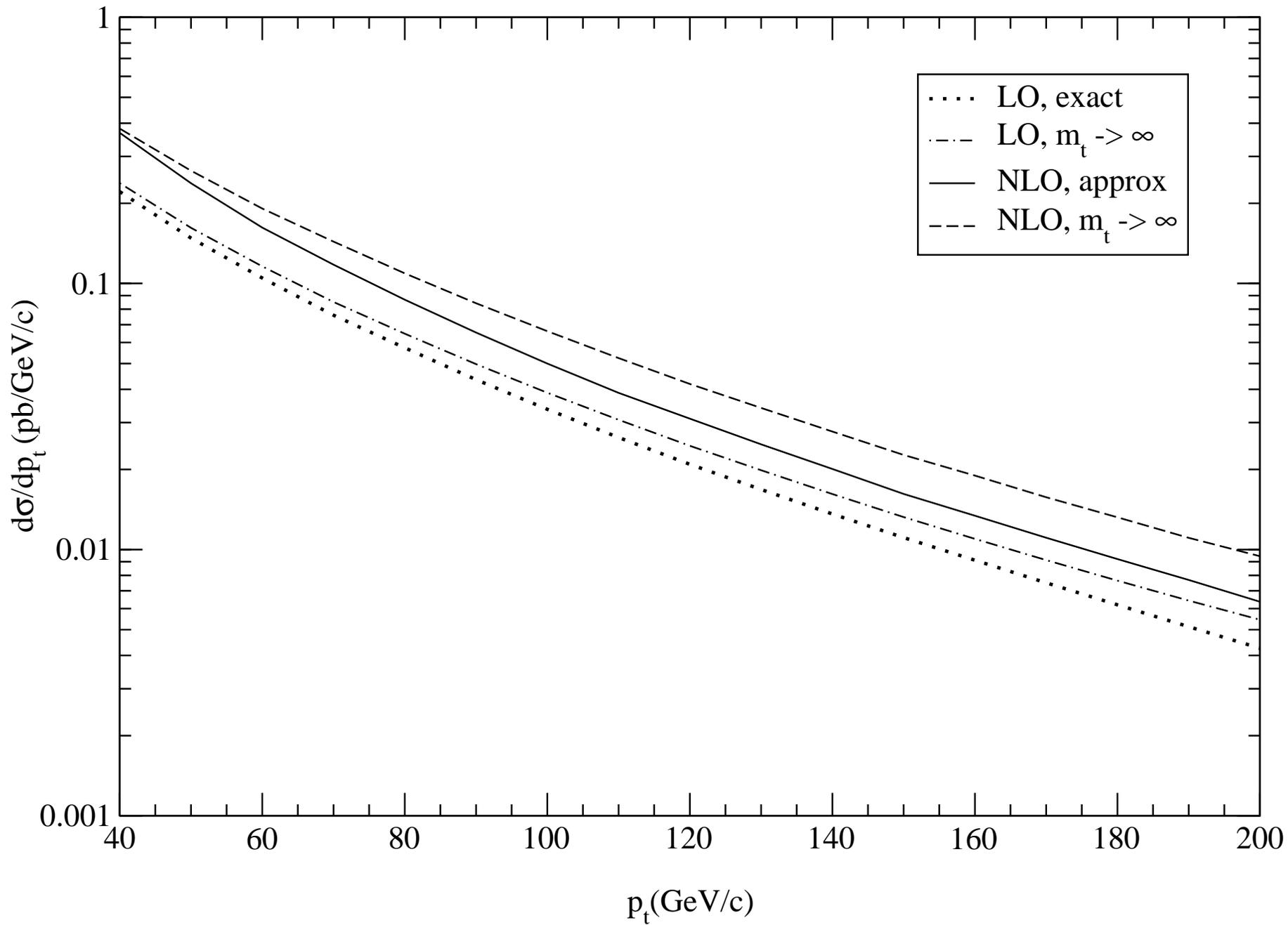
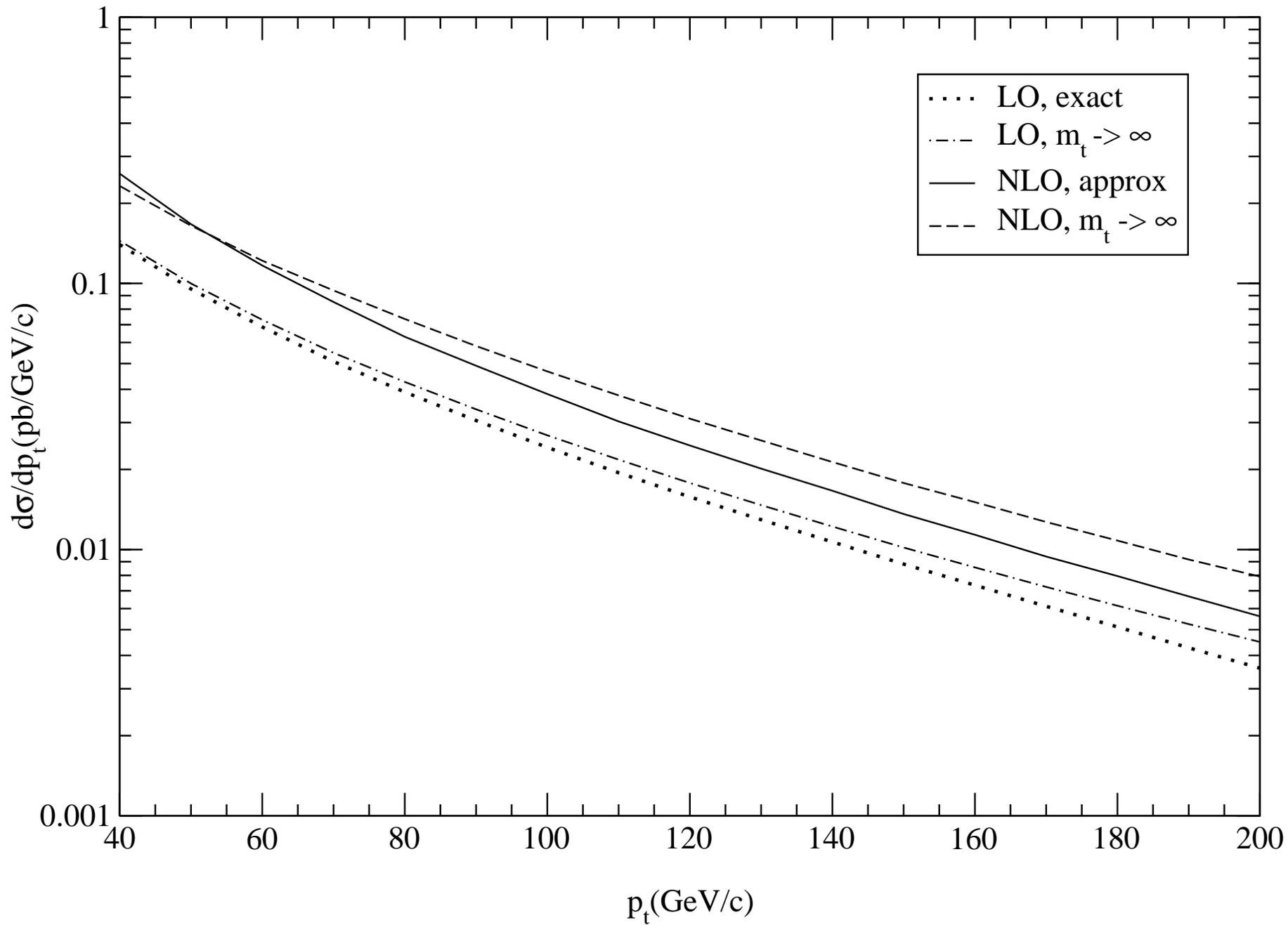
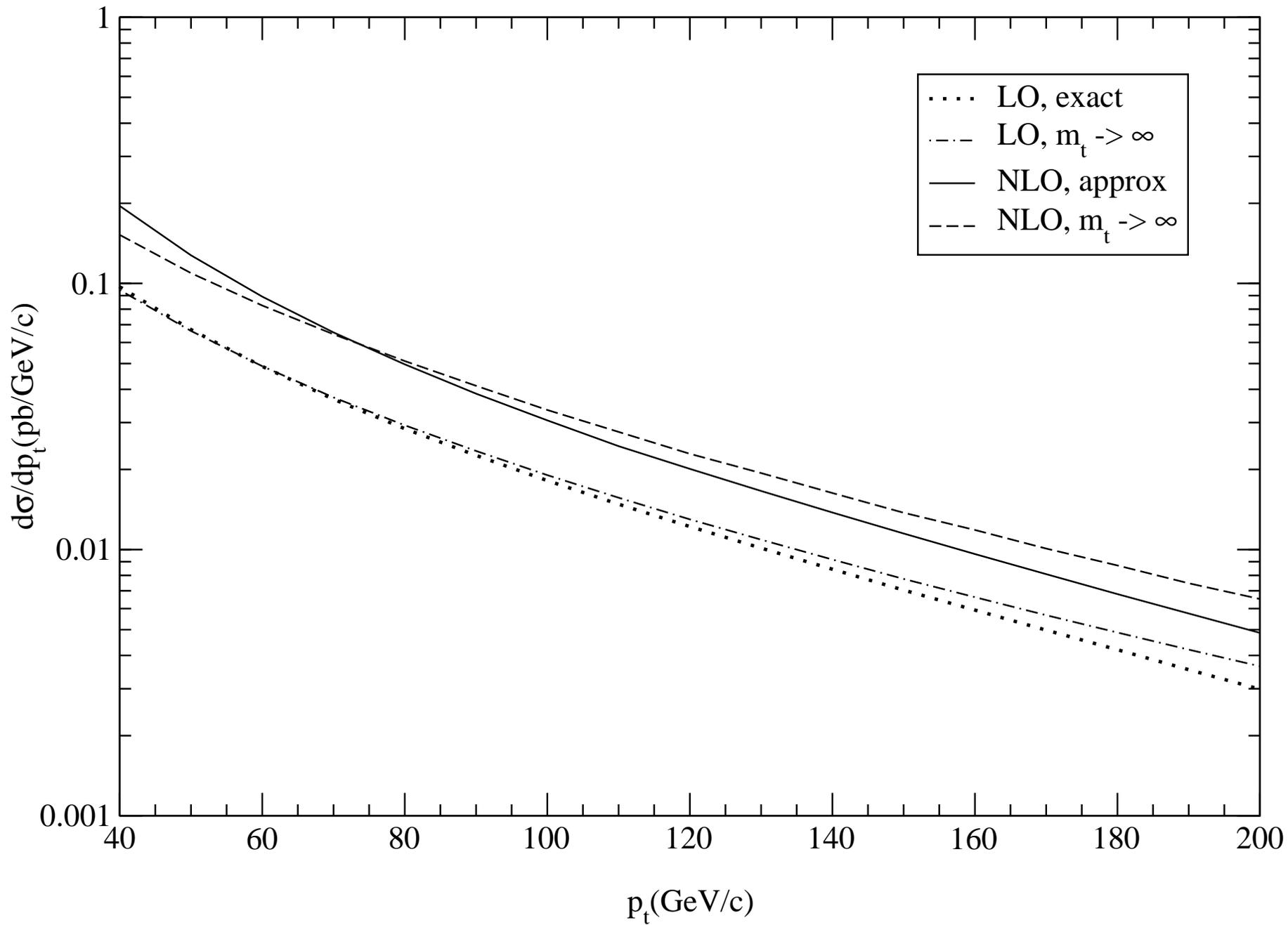
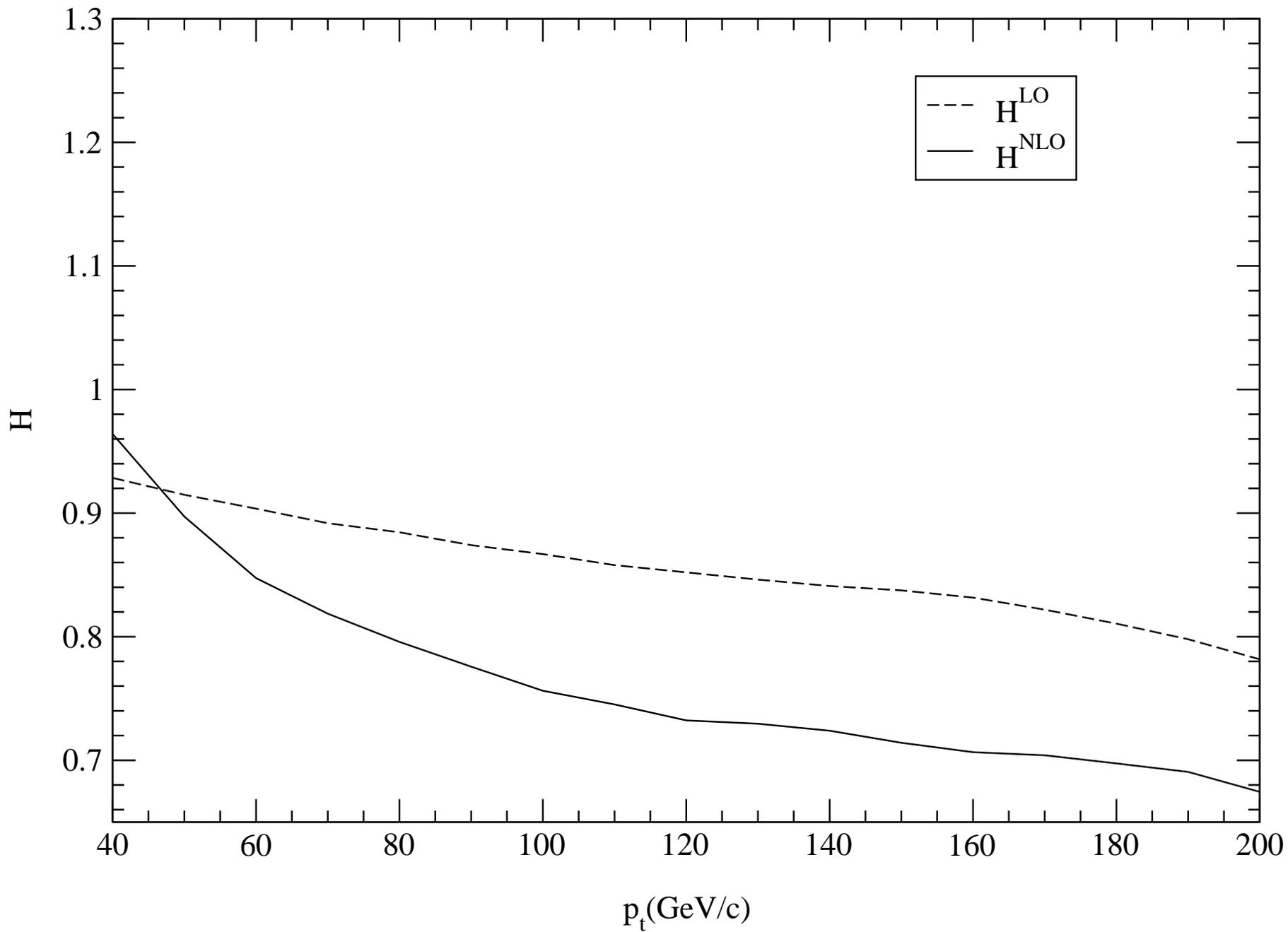
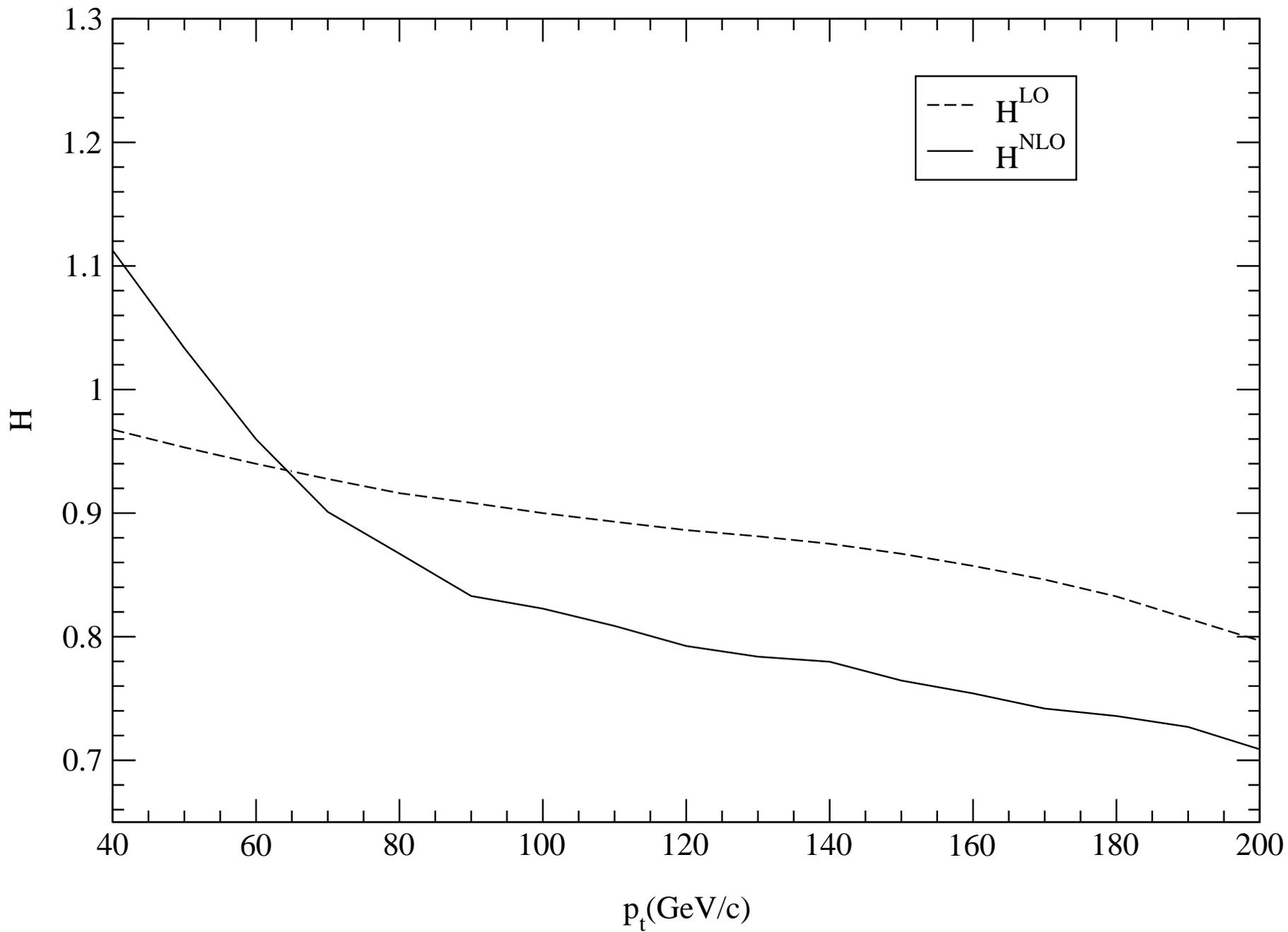
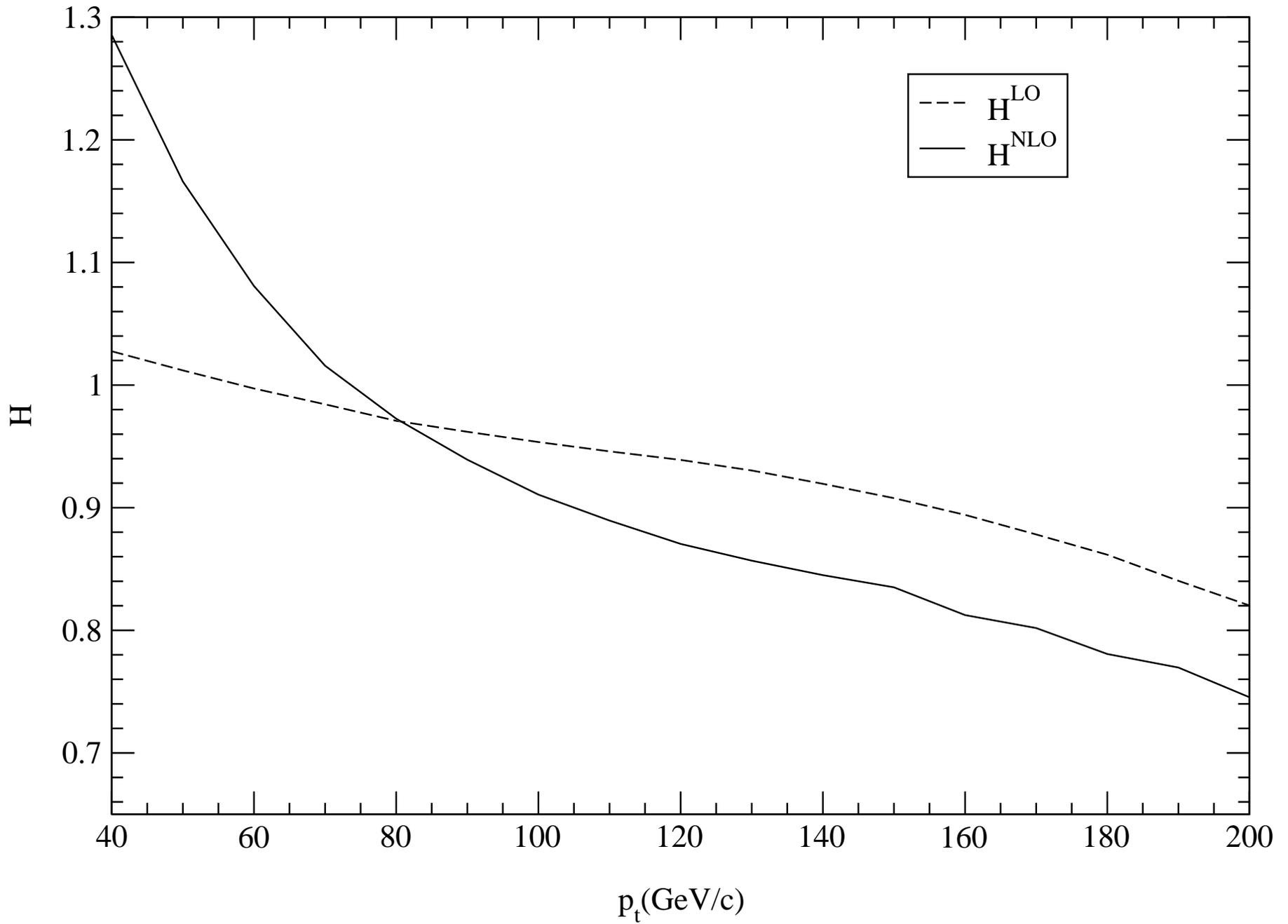
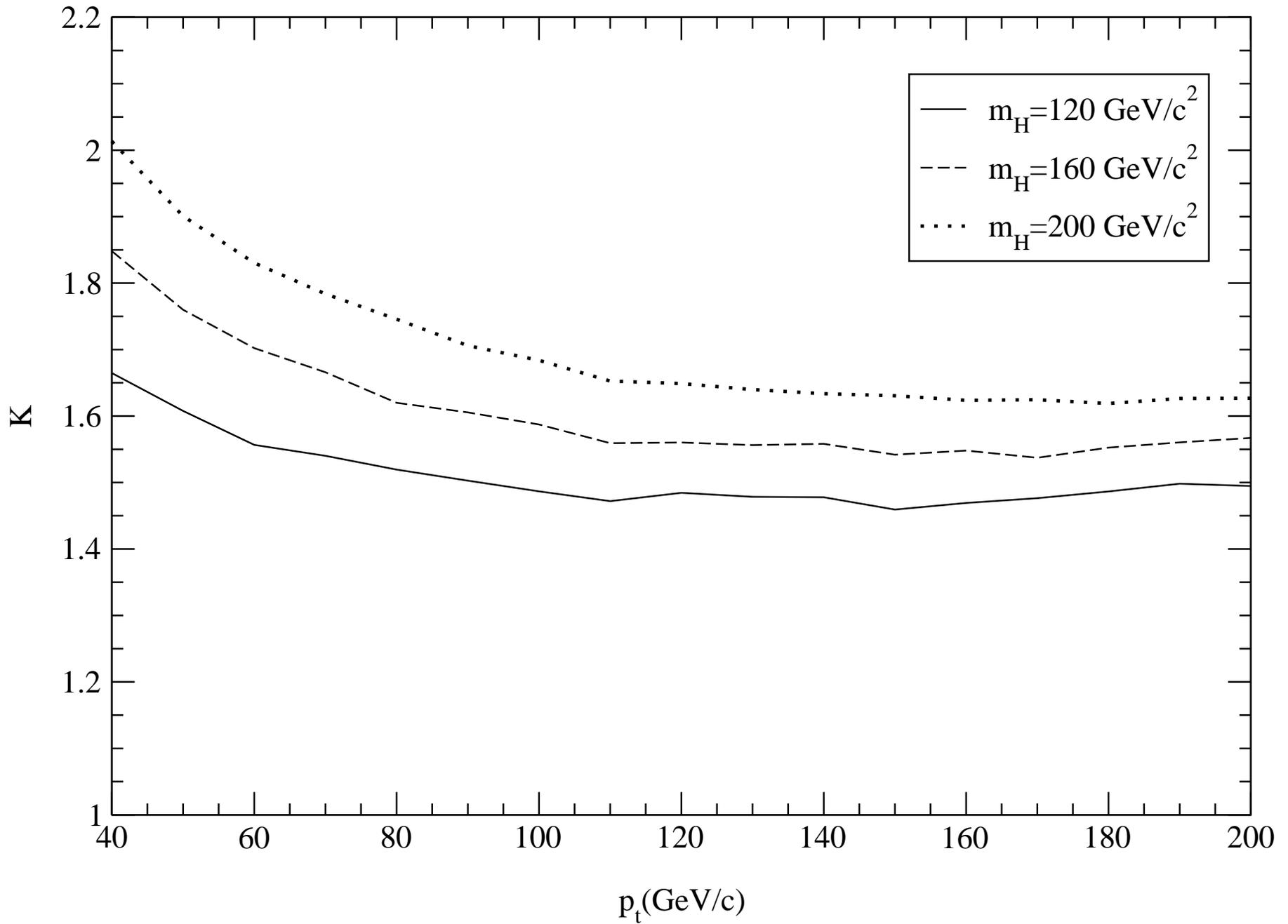
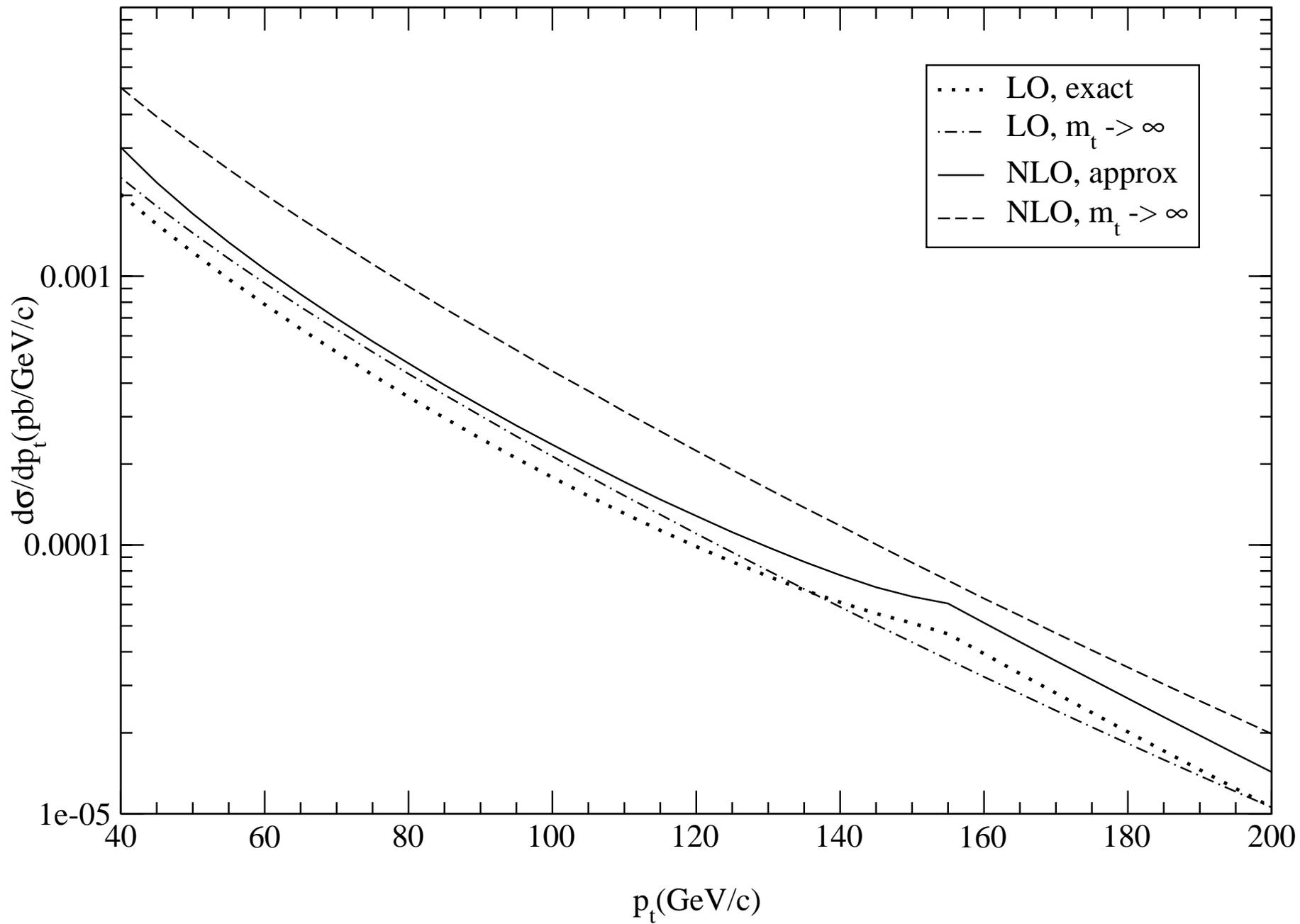
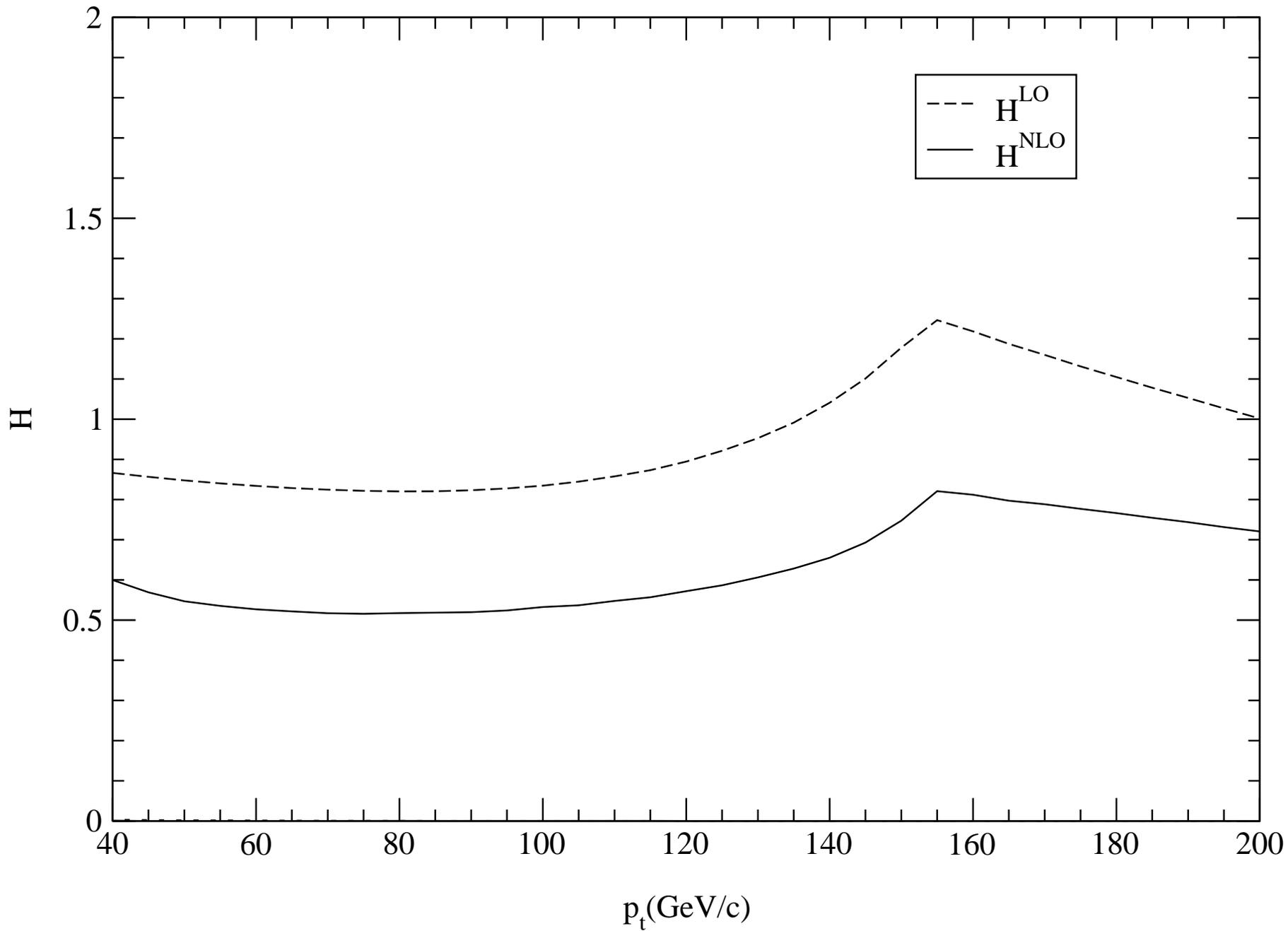
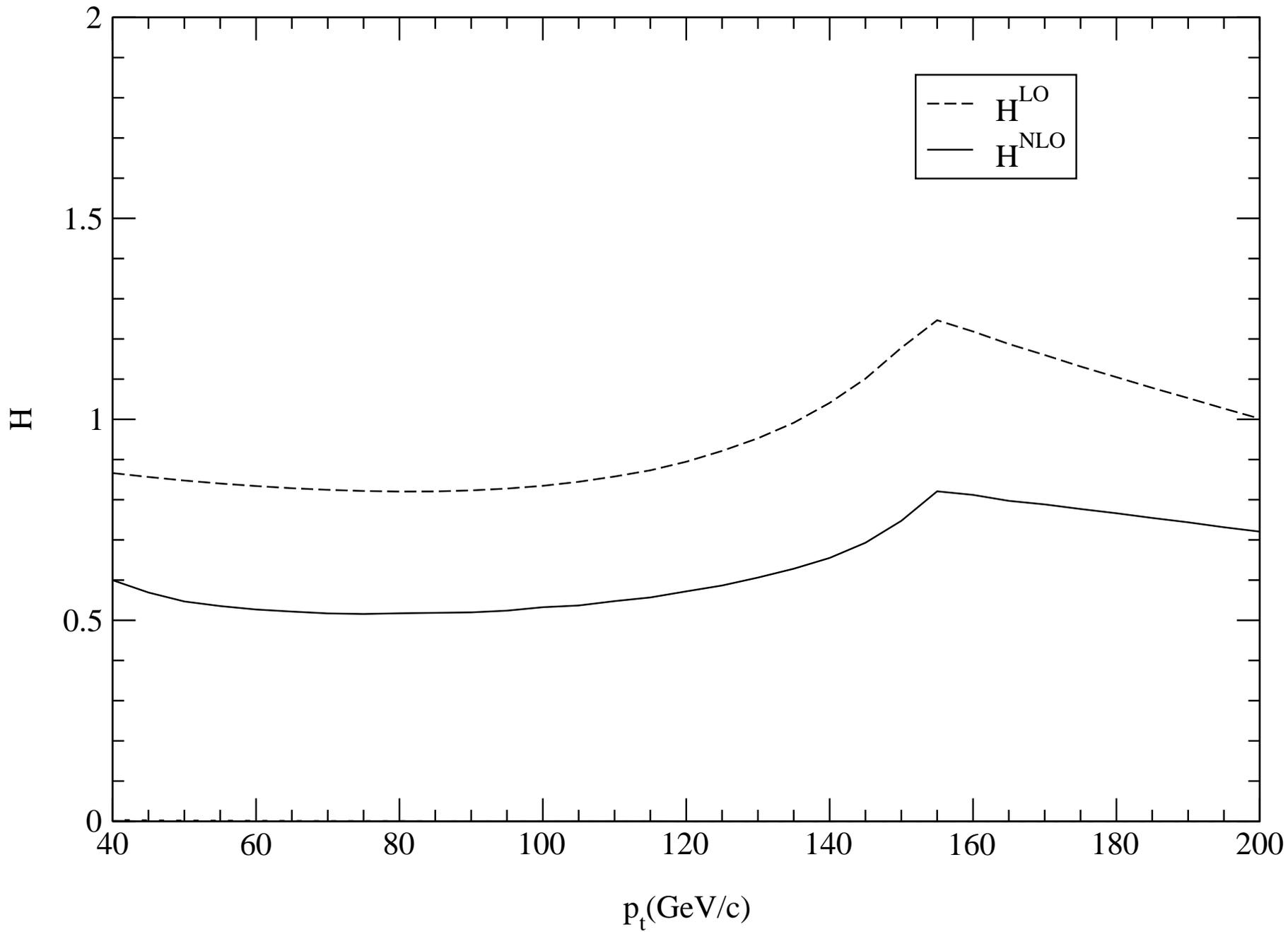
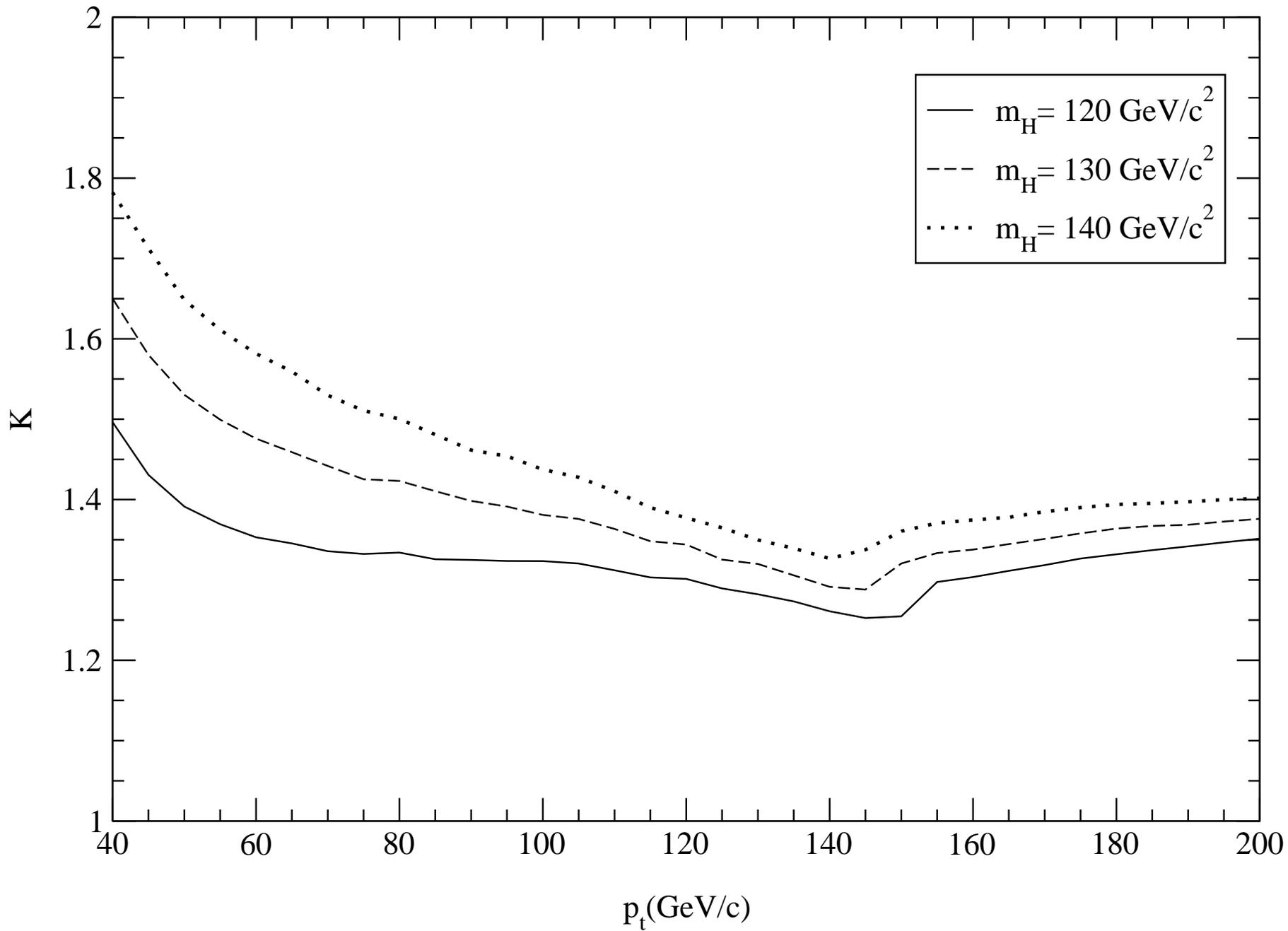
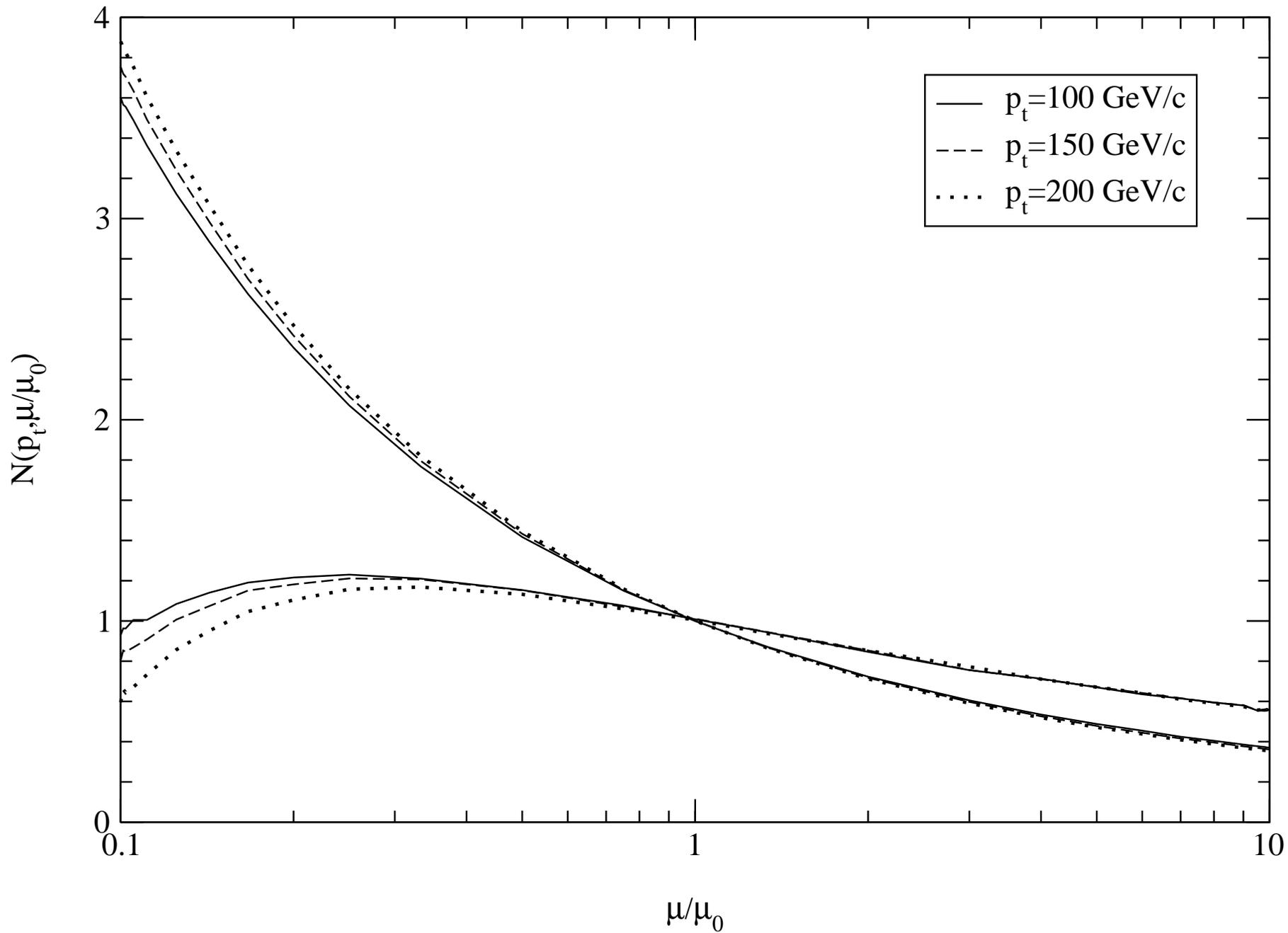
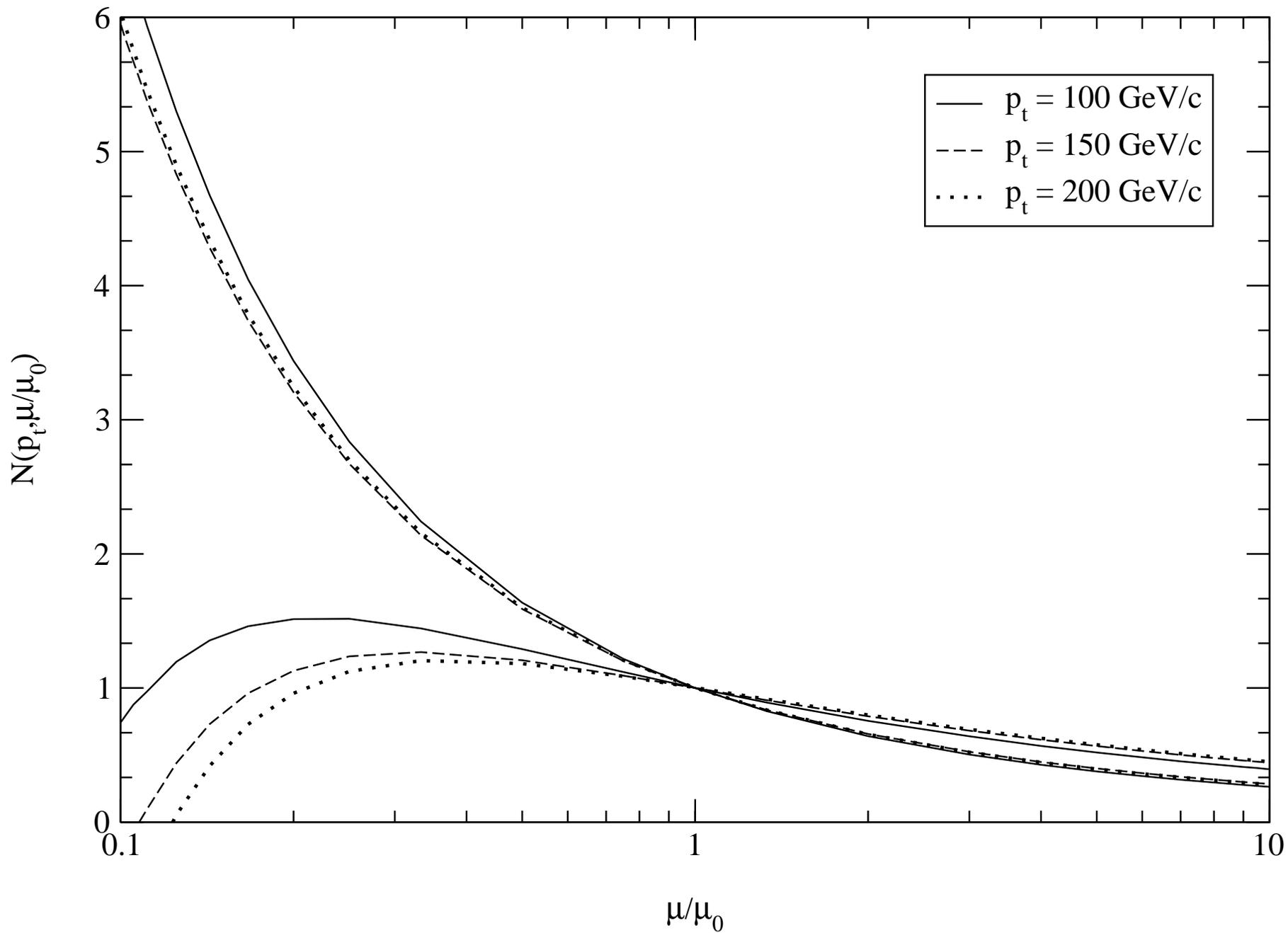
\end{document}